\pdfoutput=1
\documentclass[pdflatex, 11pt]{article}
\usepackage[english]{babel}
\usepackage{algorithm}
\usepackage{algorithmic}
\usepackage{amsfonts}
\usepackage{amsmath}
\usepackage{amssymb}
\usepackage{bbm}
\usepackage{blindtext}
\usepackage{booktabs}
\usepackage{color}
\usepackage{geometry}
\usepackage{tabularx}
\usepackage{theorem}
\usepackage{times}
\usepackage{url}
\usepackage{graphics}
\usepackage{graphicx}
\renewcommand{\baselinestretch}{1.5}

\newcommand{\Var}{\text{Var}}
\newcommand{\Cov}{\text{Cov}}

\newcommand{\eg}{\textit{e.g.}}
\newcommand{\etal}{\textit{et~al}.}

\newcommand{\vp}{\mathbf{p}}
\newcommand{\vq}{\mathbf{q}}
\newcommand{\vx}{\mathbf{x}}
\newcommand{\vy}{\mathbf{y}}
\newcommand{\vz}{\mathbf{z}}

\renewcommand{\d}{\text{d}}

\newcommand{\eps}{\epsilon}

\newcommand{\vtheta}{\text{\boldmath{$\theta$}}}

\newcommand{\DA}{\mathcal{DA}}
\newcommand{\G}{\mathcal{G}}
\newcommand{\GB}{\mathcal{GB}}
\newcommand{\SM}{\mathcal{SM}}
\newcommand{\U}{\mathcal{U}}

\begin{document}

\title{
	\textbf{Approximate Bayesian Computation for Lorenz Curves from Grouped Data}
	\thanks{
		This work is partially supported by KAKENHI (\#25245035, \#15K17036, \#16K03592, and \#16KK0081).
	}
}
\author{
	Genya Kobayashi
	\thanks{
		Graduate School of Social Sciences, Chiba University,
		Yayoi-cho 1-33, Inage-ku, Chiba 263-8522, Japan.
		TEL: +81-43-290-2405.
		\textit{Email}: \texttt{gkobayashi@chiba-u.jp}
	}
	\and
	Kazuhiko Kakamu
	\thanks{
		Institute for Economic Geography and GIScience,
		Vienna University of Economics and Business.
		Graduate School of Business Administration, Kobe University,
		2-1, Rokkodai, Nada, Kobe 657-8501, Japan.
		TEL: +81-78-803-6912.
		\textit{Email}: \texttt{kakamu@person.kobe-u.ac.jp}
	}
}
\date{%
	\today
}

\maketitle

\begin{abstract}
This paper proposes a new Bayesian approach to estimate the Gini coefficient from the Lorenz curve based on grouped data.
The proposed approach assumes a hypothetical income distribution and estimates the parameter by directly working on the likelihood function implied by the Lorenz curve of the income distribution from the grouped data.
It inherits the advantages of two existing approaches through which the Gini coefficient can be estimated more accurately and a straightforward interpretation about the underlying income distribution is provided.
Since the likelihood function is implicitly defined, the approximate Bayesian computational approach based on the sequential Monte Carlo method is adopted.
The usefulness of the proposed approach is illustrated through the simulation study and the Japanese income data.

	\vspace*{8pt}

	\vspace*{8pt}

\end{abstract}

\section{Introduction}
The Gini coefficient plays a crucial role in measuring inequality and can be a basis of political decision-making.
Although it is ideal to utilise individual household data to estimate the Gini coefficient accurately (see, e.g., Hasegawa and Kozumi, 2003), availability of individual data is usually severely limited because of the difficulty in data collection and management and confidentiality of, say, individual income data, the former being particularly true in developing countries.
Instead, grouped data, which provide the summary of income for several income classes, are widely available.
Estimating the Gini coefficient based on grouped data has drawn substantial attention from both the theoretical and empirical perspectives.
See, for example, Chotikapanich~(2008) for an overview.

There are mainly two approaches to estimate the Gini coefficient from grouped income data in the parametric framework (Ryu and Slottje,~1999).
One is to assume a hypothetical statistical distribution for income and to estimate the parameter of the distribution  (McDonald and Xu,~1995).
The Gini coefficient is then calculated from the estimates.
The advantage of this approach is that
it provides a straightforward interpretation about the underlying income distribution because the moments can be computed and the shape of the distribution can be visualised through the parameter estimates.
There exists a wide range of size distributions (Kleiber and Kotz,~2003) and the likelihood function can be constructed based on the multinomial distribution (McDonald,~1984) or selected order statistics (Nishino and Kakamu,~2011; David and Nagaraja,~2003). 
The other approach is to fit a specific functional form to the Lorenz curve and estimate the parameters of the function.
Since such a functional form for the Lorenz curve is designed so that the inequality measures can be easily derived, the Gini coefficient is immediately calculated once the parameter estimates are obtained.
A notable advantage of this approach is that a large list of functional forms is available (see, \eg, Kakwani and Podder, 1973; Basmann \etal,~1986; 1990; Ortega, \etal,~1991; Rasche \etal,~1980; Villase\~nor and Arnold,~1989; Chotikapanich,~1993; Sarabia \etal,~1999) in addition to the ones that are derived from the well-known income models such as the lognormal, Singh-Maddala, and Dagum distributions.
However, while the implied probability density function exists provided some conditions are satisfied (Iritani and Kuga,~1983 and Sarabia,~2008), the interpretation as a statistical size distribution is less intuitive as the support of the probability density function of the implied distribution is limited to some interval.
Moreover, current practice for parametric Lorenz curve estimation lacks a solid statistical foundation compared with hypothetical statistical distribution estimation despite the fact that the discussions of the Lorenz curve have long history since the seminal work by Lorenz (1905).

Given this context, the aim of the current paper is to estimate the Lorenz curve of a hypothetical statistical distribution from the grouped data in a general framework. 
For some flexible hypothetical distribution, such as the generalised beta distribution, an analytical form of the Lorenz curve is not available and the evaluation of the Lorenz curve can be computationally expensive and unstable. 
Therefore, a new estimation procedure for the case where the Leorenz curve is not available explicitly is required.
Moreover, it is worth noting that the present study is motivated by the approach proposed by Chotikapanich and Griffiths~(2002, 2005). 
More recently, Hajargasht and Griffiths (2015) proposed a generalised method-of-moment approach for the Lorenz curves in a similar setting.
In Chotikapanich and Griffiths~(2002, 2005), the expectation of the income shares for the groups, and hence that of the differences between the Lorenz curves for the consecutive groups, is assumed to be equal to the population shares for the corresponding groups.
Then, they adopted the likelihood function based on the Dirichlet distribution and proposed the maximum likelihood estimator and Bayes estimator by using the Markov chain Monte Carlo (MCMC) method.
Although their Dirichlet likelihood approach may be convenient, the parameter estimates and the resulting Gini estimate can be highly sensitive with respective to the additional tuning parameter, which is required to construct the Dirichlet likelihood.
In the Bayesian framework, the posterior distributions of the parameters and Gini coefficient are sensitive with respect to the prior distribution of this parameter.
Furthermore, the evaluation of the likelihood function requires the evaluation of the Lorenz curve for each group.
The applicability of this approach is limited to cases where evaluation of the Lorenz curve derived from the hypothetical distribution is feasible or where substantial prior information for the tuning parameter is available.

Motivated by the above issues, in this paper, we attempt to work on the likelihood function implied from the Lorenz curve of the hypothetical income distribution, instead of working on the Dirichlet likelihood.
We employ the approximate Bayesian computation (ABC) method that avoids the direct evaluation of the likelihood function and simulate data from the model, given a candidate parameter value.
If the simulated and observed data are similar, the candidate parameter value is a good candidate to generate the observed data.
Then, it can be regarded as a sample from the posterior distribution.
The ABC method has a wide variety of fields of application including population genetics, population biology, signal processing, epidemiology, and economics.
See, for example, Csill\'ery \etal~(2010), Sisson and Fan~(2011), and Marin~\etal~(2012) for an overview of the ABC methods.
The application of the ABC method requires an ability to simulate datasets from the probability model and is well-suited to the present context, because datasets can be easily generated from the hypothetical income distribution.
Since it is difficult to devise an efficient proposal distribution for an MCMC algorithm in the ABC setting, we adopt the sequential Monte Carlo (SMC) algorithm with adaptive weights proposed by Bonassi and West~(2015), which is computationally efficient and easy to implement.

The rest of this paper is organised as follows.
Section~\ref{sec:sec2} briefly reviews the estimation methods for the Lorenz curve from grouped data
and proposes our estimation method based on ABC.
The five-parameter generalised beta distribution is adopted as a flexible hypothetical income distribution.
Section~\ref{sec:ne} illustrates the proposed method by using the simulated data and compares the performance with the existing methods.
The application of the proposed method to the real data from the Family Income and Expenditure Survey in Japan is also presented.
Finally, Section~\ref{sec:conc} concludes and some remaining issues are discussed.

\section{Method}\label{sec:sec2}
\subsection{Estimating Gini Coefficient from Lorenz Curve Based on Grouped Data}\label{sec:sec21}
Suppose that the population is divided into $k$ groups.
Let us denote the observed cumulative share of households and income by $\vp = (p_{0} = 0, p_{1}, \dots, p_{k-1}, p_{k} = 1)$ and $\vy = (y_{0} = 0, y_{1}, \dots, y_{k-1}, y_{k} = 1)$, respectively, which are usually constructed from a survey on $n$ individual households.
Even if the cumulative share of households and cumulative share of income are not directly available, 
$\vp$ and $\vy$ can be calculated from the income classes, class income means, and number of households reported in the grouped data. 
Let us denote the cumulative distribution function and probability density function of the hypothetical income distribution with the parameter $\vtheta$ by $F(\cdot|\vtheta)$ and $f(\cdot|\vtheta)$, respectively.
Then, the Lorenz curve denoted by $L(y|\vtheta)$ is defined by
\[
	L(y|\vtheta) = \frac{1}{\mu} \int_{0}^{y} F^{-1}(z) \d z, \quad y \in [0,1],
\]
where $\mu$ is the mean of the distribution and $F^{-1}(z) = \inf \left\{ x: F(x) \ge z \right\}$.
Once the parameter estimate for $\vtheta$ is obtained, the Gini coefficient can be estimated by using
\begin{eqnarray}
	G &=& -1 +\frac{2}{\mu} \int_{0}^{\infty} x F(x) f(x) \d x,
	\label{eqn:gini1}\\
  &=& 1 - 2 \int_{0}^{1} L(z) \d z,
	\label{eqn:gini2}
\end{eqnarray}

There are several methods to estimate the parameters of the Lorenz curves, for example, the least squares (Kakwani and Podder,~1973) or generalised least squares (Kakwani and Podder,~1976).
More recently, Chotikapanich and Griffiths~(2002) proposed a maximum likelihood estimator based on the likelihood from the Dirichlet distribution given by
\begin{equation}\label{eqn:cg}
	f(\vq| \vtheta,\lambda) = \Gamma(\lambda) \prod_{j = 1}^{k}\frac{q_{j}^{\lambda(L(p_{j}|\vtheta) - L(p_{j-1}|\vtheta))- 1}}{\Gamma(\lambda(L(p_{j}|\vtheta) - L(p_{j-1}|\vtheta)))},
\end{equation}
where $\vq = (q_{1}, \dots, q_{k}$),  $q_{i} = y_{j} - y_{j-1}$ is the income share for the $j$-th group, $\Gamma(\cdot)$ is the gamma function, and $\lambda$ is the additional parameter of the Dirichlet likelihood.
This likelihood function is motivated by the assumption given by $E[q_{j}] = L(p_{j}|\vtheta)-L(p_{j-1}|\vtheta)$.
The variance and covariance of the income share implied from this likelihood are given by
\[
	\Var(q_{j})=\frac{E[q_{j}](1 - E[q_{j}])}{\lambda + 1}, \quad \Cov(q_{i}, q_{j}) = -\frac{E[q_{i}]E[q_{j}]}{\lambda + 1},
\]
where $\lambda$ acts as a precision parameter.
Larger values of $\lambda$ suggest that the smaller variation of the income shares are implied from the Lorenz curve.
Based on this likelihood function, Chotikapanich and Griffiths~(2005) considered an MCMC method in the Bayesian framework by specifying the prior distributions of $\vtheta$ and $\lambda$.

Although their Dirichlet likelihood approach may appear convenient, it has the following problems.
The parameter estimates and the resulting Gini estimate can be highly sensitive with respective to the choice of the value or prior distribution of $\lambda$, since data do not contain information on this parameter.
The sensitivity is especially profound when the number of groups is small.
Furthermore, the evaluation of the likelihood function requires the evaluation of $L(y_{i}|\vtheta)$ for $i = 1, \dots, k$.
Except for some simple standard distributions, such as the lognormal, Singh-Maddala, and Dagum, some flexible classes of hypothetical income distributions do not admit an analytical form of the Lorenz curve or the evaluation of the Lorenz curve is computationally expensive and unstable.
Therefore, the inference based on the Dirichlet approach can be unreliable and its applicability would be limited.

\subsection{Hypothetical Income Distribution: Generalised Beta Distribution}
In order to estimate the Lorenz curve and the related inequality measures accurately, a flexible class of hypothetical distributions is required.
This paper adopts the five-parameter generalised beta (GB) distribution denoted by $\GB(a,b,c,p,q)$ as an interesting and important income distribution.
This distribution was by proposed by McDonald and Xu~(1995) and is the most flexible distribution of the family of beta-type distributions.
The probability density function of the GB distribution is given by
\begin{eqnarray}
	f_{GB}(x) = \frac{\displaystyle |a| x ^ {a p - 1}\left[ 1 - (1 - c) \left( \frac{x}{b} \right)^{a} \right]^{q - 1}}{\displaystyle b ^ {a p} B(p, q) \left[ 1 + c \left( \frac{x}{b} \right)^{a} \right]^{p + q}},\quad 0 < x^{a} < \frac{b^{a}}{1 - c},
	\label{GB:prob}
\end{eqnarray}
where $a\in\mathbb{R}$, $b>0$, $c\in[0,1]$, $p>0$, $q>0$, and $B(p, q)$ is the beta function.
Using the incomplete beta function $B_x(p,q)$, the cumulative distribution function is given by $F_{GB}(x)=B_z(p,q)/B(p,q)$ with $z=(x/b)^a/(1+c(x/b)^a)$.
Given a set of parameter values, the Gini coefficient can be easily computed from \eqref{eqn:gini1} by using numerical integration.
The GB distribution includes a number of special cases.
For example, when $c=0$ and $c=1$, the GB distribution reduces to the generalised beta distribution of the first and second kind (GB1 and GB2) (McDonald,~1984), respectively.
Moreover, when $(c,p)=(1,1)$ and $(c,q)=(1,1)$, the distribution reduces to the Singh-Maddala (SM) distribution (Singh and Maddala,~1976) and Dagum (DA) distribution (Dagum,~1977), which are known to perform well in many empirical applications.
Detailed relationships among the class of distributions are summarised in McDonald and Xu~(1995).

The hypothetical GB distribution can also be directly estimated from the grouped level income data by using the MCMC or maximum likelihood method (Kakamu and Nishino,~2016). 
An explicit formula of the Lorenz curve for the GB distribution is not available (McDonald and Ransom, 2008). 
This is also the case for the GB2 distribution, although the result on the Lorenz ordering for GB2 is known (Sarabia~\etal,~2002).
Hence, the likelihood  based on the Dirichlet distribution \eqref{eqn:cg} is not explicitly available and the evaluation of the likelihood can be computationally expensive and unstable. 
Note that the random variable $X\sim\GB(a,b,c,p,q)$ can be easily generated by using
\begin{equation}\label{eqn:rangb}
X = b \left( \frac{Z}{1 - c Z} \right)^{\frac{1}{a}},
\end{equation}
where $Z\sim Be(p,q)$ (Kakamu and Nishino,~2016).
Since the Lorenz curve is location-free, we let $\vtheta=(a,c,p,q)'$ and $b$ is fixed to $1$.
Therefore, the proposed ABC method described in the following  would be a convenient approach to estimating the hypothetical GB distribution from the Lorenz curve based on the grouped data.

\subsection{Approximate Bayesian Computation for Lorenz Curve}\label{sec:abc}
We work on the likelihood function implied from the Lorenz curve of the hypothetical income distribution.
This likelihood function is constructed through the statistics of the individual household incomes whose distribution is not explicitly available.
Thus, the standard MCMC methods cannot be directly applied, because these methods require evaluating the likelihood function and prior density.
The approximate Bayesian computation (ABC) methods avoid  direct evaluation of the likelihood function and simulate data from the model, given a candidate parameter value.
If the simulated and observed data are similar, the candidate parameter value is a good candidate to generate the observed data.
Then, it can be regarded as a sample from the posterior distribution (Sisson and Fan,~2011).
The posterior distribution can be approximated by weighting the intractable likelihood function.
Therefore, ABC is a convenient approach when the likelihood function is not explicitly available or computationally prohibitive to evaluate, because it requires only being able to simulate data from the probability model.

Let $\pi(\vtheta)$ denote the prior density of the parameter $\vtheta$, $f(\vy|\vtheta)$ be the likelihood function of the observed data $\vy$, and $\pi(\vtheta|\vy)\propto f(\vy|\vtheta)\pi(\vtheta)$ be the posterior distribution of $\vtheta$.
ABC methods augment the posterior from $\pi(\vtheta|\vy)$ to
\[
\pi_\eps(\vtheta,\vx|\vy)\propto \pi(\vtheta)f(\vx|\vtheta) I_{A_{\eps,\vy}}(\vx),
\]
where $\eps>0$ is a tolerance level, $I_B(\cdot)$ is the indicator function of the set $B$, and $\vx$ is the simulated data.
The set $A_{\eps,\vy}$ is defined as
$A_{\eps,\vy}=\left\{\vx: \rho(\vx,\vy)<\eps\right\}$,
where $\rho(\cdot,\cdot)$ is a distance function.
The value of $\eps$ and form of $\rho$ are chosen by the user and can affect the performance of ABC.
The marginal distribution
\[
\pi_\eps(\vtheta|\vy)\propto\int  \pi(\vtheta)f(\vx|\vtheta) I_{A_{\eps,\vy}}(\vx)\d\vx
\]
provides an approximation to $\pi(\vtheta|\vy)$ for sufficiently small $\eps$.

Various ABC algorithms to sample from the approximate posterior distribution based on, for example, rejection sampling (Beaumont \etal,~2002), MCMC (Marjoram \etal,~2003; Fearnhead and Prangle,~2012), and the sequential Monte Carlo (SMC) method (Sisson \etal~2007,~2009; Beaumont \etal,~2009; Toni \etal,~2009) have been proposed.
Furthermore, a number of extensions of the SMC algorithm has been considered by, for example, Del~Moral \etal~(2012), Lenormand \etal~(2013), Filippi \etal~(2013), Silk \etal~(2013), and Bonassi and West~(2015).
We employ the SMC approach because it is difficult to construct an efficient proposal distribution for MCMC in the present context and the SMC with adaptive weights proposed by Bonassi and West~(2015), among others, is adopted because of its computational efficiency and ease of implementation.
The SMC algorithm proceeds by sampling from a series of intermediate distributions with user-specified decreasing tolerance levels, $\pi_{\eps_t}(\vtheta,\vx|\vy)$ with $\eps_t<\eps_{t-1}$ for $t=0,\dots,T$ and $\eps_T$.
A large number of particles, denoted by $(\vtheta_i,\vx_i)$, $i=1,\dots,N$, is propagated by using importance sampling and resampling until the target tolerance level $\eps_T$ is reached.
Bonassi and West~(2015) proposed to approximate the intermediate distribution at each step by kernel smoothing with the joint kernel $K_t(\vtheta,\vx|\tilde{\vtheta},\tilde{\vx})$.
Bonassi and West~(2015) employed the product kernel such that $K_t(\vtheta,\vx|\tilde{\vtheta},\tilde{\vx})=K_{t,\vtheta}(\vtheta|\tilde{\vtheta})K_{t,\vx}(\vx|\tilde{\vx})$.
Note that $K_{t,\vx}(\vx|\tilde{\vx})$ is uniform over $A_{\eps_t,\vy}$ in the standard SMC of Sission,~(2007, 2009), Beaumont \etal~(2009), and Toni \etal~(2009).
Algorithm~\ref{alg} describes the method of Bonassi and West~(2015).
Introducing a kernel function for $\vx$ makes the perturbation step of the algorithm such that particles for which the simulated $\vx$ is close to $\vy$ are  chosen more likely.
Bonassi and West~(2015) showed that the proposal distribution of their algorithm has higher prior predictive density over the acceptance region for the next step---and hence, higher acceptance probability---than that of the standard SMC algorithm.
Finally, the posterior distribution of $\vtheta$ at step $t$ of the algorithm is approximated by
\[
\pi_{\eps_t}(\vtheta|\vy)\propto \int \int \pi(\tilde{\vtheta})f(\tilde{\vx}|\tilde{\vtheta})K_{t,\vtheta}(\vtheta|\tilde{\vtheta}) K_{t,\vx}(\vy|\tilde{\vx})I_{A_{\eps_t,\vy}}(\vx) \d \tilde{\vx} \d\tilde{\vtheta}.
\]

\renewcommand{\baselinestretch}{1.0}
\begin{algorithm}[h]
\caption{\textit{SMC with adaptive weights}
\label{alg}
}
\medskip
\begin{algorithmic}[1]
\setlength{\itemsep}{3pt}
\setlength{\parsep}{0pt}
\STATE
Initialise tolerance levels $\eps_0>\eps_1>\dots>\eps_T$ and set $t=0$.
\FOR{$i=1$ to $N$}
\REPEAT
\STATE
Simulate $\vtheta_i^{(0)}$ from $\pi(\vtheta)$ and $\vx_i^*$ from $f(\vx|\vtheta_i^{(0)})$.
\UNTIL{$\rho(\vx_i^*,\vy)<\eps_0$}
\STATE
Set the weights $w_i=1/N$ for $i=1,\dots,N$.
\ENDFOR

\FOR{$t=1$ to $T$}
\STATE
Compute the weights $v_i^{(t-1)}\propto w_i^{(t-1)}K_{\vy,t}(\vy|\vx_i)$ for $i=1,\dots,N$.
\FOR{$i=1$ to $N$}\label{alg:para1}
\REPEAT\label{alg:repeat}
\STATE
Choose $\vtheta_i^*$ from $\{\vtheta_i^{(t-1)}\}$ based on the weights $\{v^{(t-1)}_j\}$.
\STATE
Draw $\vtheta_i^{(t)}$ from $K_{\vtheta,t}(\vtheta_i^{(t)}|\vtheta_i^*)$ and simulate $\vx_i$ from $f(\vx|\vtheta_i^{(t)})$.
\UNTIL{$\rho(\vx_i,\vy)<\eps_t$}\label{alg:until}
\STATE
Compute the new weights as $w_i\propto \frac{\pi(\vtheta_i^{(t)})}{\sum_{j=1}^Nv_j K_{\vtheta,t}(\vtheta_i^{(t)}|\vtheta_j^{(t-1)})}$.
\ENDFOR\label{alg:para2}
\ENDFOR
\end{algorithmic}
\end{algorithm}
\renewcommand{\baselinestretch}{1.5}

To estimate the hypothetical income distribution from the Lorenz curve based on the group data  by using ABC, we use  the cumulative income shares in percentage.
In Algorithm~\ref{alg}, we set $\rho(\vx,\vy)=\max_j|x_j-y_j|$, which was also employed in McVinish~(2012), because the choice of the tolerance schedule and level of approximation is intuitive.
To simulate $\vx$, $n$ observations independently and identically distributed to the hypothetical income distribution is generated.
Then, they are sorted in the ascending order, denoted by $(z_{(1)},\dots,z_{(n)})$, and the cumulative income share is computed using $x_j=\sum_{i=1}^{n_j} z_{(i)}/\sum_{i=1}^n z_{(i)}$, where $n_j=\lfloor n p_j\rfloor$ for $j=1,\dots,k-1$.
For the GB distribution and its special cases, the simulated data are generated by using \eqref{eqn:rangb}. 

As in Bonassi and West~(2015), the product of normal kernels is used for $\vtheta$.
Following the rule of thumb for the product of normal kernels, the bandwidth is determined based on $h_s=\hat{\sigma}_sN^{-1/(d+4)}$, where $N$ is the number of particles, $d$ is the total dimension of the parameter and data, and  $\hat{\sigma}_s$ is the standard deviation for $s\in\{\vtheta,\vx\}$ (Scott and Sain,~2005; Bonassi and West,~2015).
When the number of groups is large, such as $k=10$ in decile data, the performance and computing time of ABC may be affected (see, \eg, Prangle,~2015), as Algorithm~\ref{alg} compares two nine-dimensional vectors.
To reduce the dimensionality, we can also use summary statistics that consist of a subset of the elements of the cumulative incomes.
For example, when $k=10$, we can replace $\vy$ and $\vx$ in Algorithm~\ref{alg} with
$S(\vy)=(y_1,y_3,y_5,y_7,y_9)$ and $S(\vx)=(x_1,x_3,x_5,x_7,x_9)$, respectively.
Note that if we take $S(\vx)=(x_2,x_4,x_6,x_8)$, it is identical to the simulated data in the case of $k=5$.
The use of the summary statistics in the case of $k=10$ is also examined in Section~\ref{sec:ne}.

\section{Numerical Examples}\label{sec:ne}
\subsection{Simulated Data 1}\label{sec:sim1}
A series of simulation studies is conducted to illustrate the proposed approach, which is denoted by ABC hereafter. 
First, the individual household income follows the Dagum (DA) and Singh-Maddala (SM) distributions, denoted by $\DA(a,b,p)=\GB(a,b,1,p,1)$ and $\SM(a,b,q)=\GB(a,b,1,1,q)$, respectively.
The performance of ABC is compared with that of the two existing methods.
The first method is based on the Dirichlet likelihood given by \eqref{eqn:cg}, since both distributions allow the explicit forms of Lorenz curves. 
The parameters $(a,p,\lambda)$ and $(a,q,\lambda)$ are estimated by using the Metropolis-Hastings (MH) algorithm. 
This approach is denoted by Dirichlet hereafter. 
The other method, proposed by Kakamu and Nishino~(2016), estimates the hypothetical income distribution from the grouped level income. 
The likelihood function is constructed from the selected order statistics (SOS), $\vz=(z_1,\dots,z_{k-1})$, given by
\begin{equation*}
f(\vz|\vtheta)=n!\frac{F(z_1|\vtheta)^{{n_1}-1}}{(n_1-1)!}f(z_1|\vtheta)\left[\prod_{j=2}^{k-1}\frac{(F(z_j|\vtheta)-F(z_{j-1}|\vtheta))^{n_j-n_{j-1}-1}}{(n_j-n_{j-1}-1)!}\right]\frac{(1-F(z_k-1))^{n-n_{k-1}}}{(n-n_{k-1})!},
\end{equation*}
where $n_j,\ j=1,\dots,k-1$ is defined in Section~\ref{sec:abc}.
This likelihood is similar to the multinomial likelihood, but it provides a more accurate representation for the grouped data, (David and Nagaraja,~2003). 
The posterior inference is based on the MH algorithm. 
This approach is denoted by SOS hereafter.

To create the data for this simulation study, $n=10000$ observations are generated from $\DA(a,1,p)$ and $\SM(a,1,q)$.
Then, the data are sorted in ascending order and are grouped into $k$ groups to calculate the income and household shares.
The data are replicated 100 times.
For the Dagum distribution, the following four settings for the true parameter values and corresponding Gini coefficients are considered:
(i) $(a,p,G)=(3.8,1.3,0.2482)$,
(ii) $(a,p,G)=(3.0,1.5,0.3087)$,
(iii)$(a,p,G)=(2.5,2.5,0.3518)$,
(iv)$(a,p,G)=(2.3,1.5,0.4077)$.
For the Singh-Maddala distribution, the following four settings are considered:
(i) $(a,q,G)=(3.5,1.5,0.2429)$,
(ii) $(a,q,G)=(2.3,3.0,0.3041)$,
(iii)$(a,q,G)=(2.0,2.5,0.3567)$,
(iv)$(a,q,G)=(1.6,3.5,0.4052)$.
For the number of groups, we consider $k=5$ and $10$.
These choices respectively correspond to quintile and decile data, which are the most commonly available in practice.
In the case of $k=10$, we also implement Algorithm~\ref{alg} with $S(\vy)=(y_1,y_3,y_5,y_7,y_9)$ and $S(\vx)=(x_1,x_3,x_5,x_7,x_9)$ suggested in Section~\ref{sec:abc}.

For ABC, we used 3000 particles with the schedule of tolerance levels given by $\{\eps_t\}=\left\{0.1,0.01, 0.002\right\}$.
Algorithm~\ref{alg} is implemented by using Ox Professional version 7.10 (Doornik,~2013) with six parallel threads for the lines between \ref{alg:para1} and \ref{alg:para2}.
For all the methods, we assume $a\sim\G(3,1)$, $b\sim\G(3,1)$, $p\sim\G(3,1)$, $q\sim\G(3,1)$ to reflect the results in the existing literature on the GB distributions (\eg, McDonald and Ransom,~2008).
To illustrate the prior sensitivity of Dirichlet, the following prior distributions of $\lambda$ are considered: $\G(1,0.1)$, $\G(1,0.5)$, and $\G(1,1)$.
For SOS and Dirichlet, the MCMC algorithms are run for $40000$ iterations including the $10000$ initial burn-in period.
To reduce the undesired autocorrelation among the MCMC samples, every 10th draw is retained for posterior inference.

\begin{figure}[H]
\centering
\includegraphics[width=\textwidth]{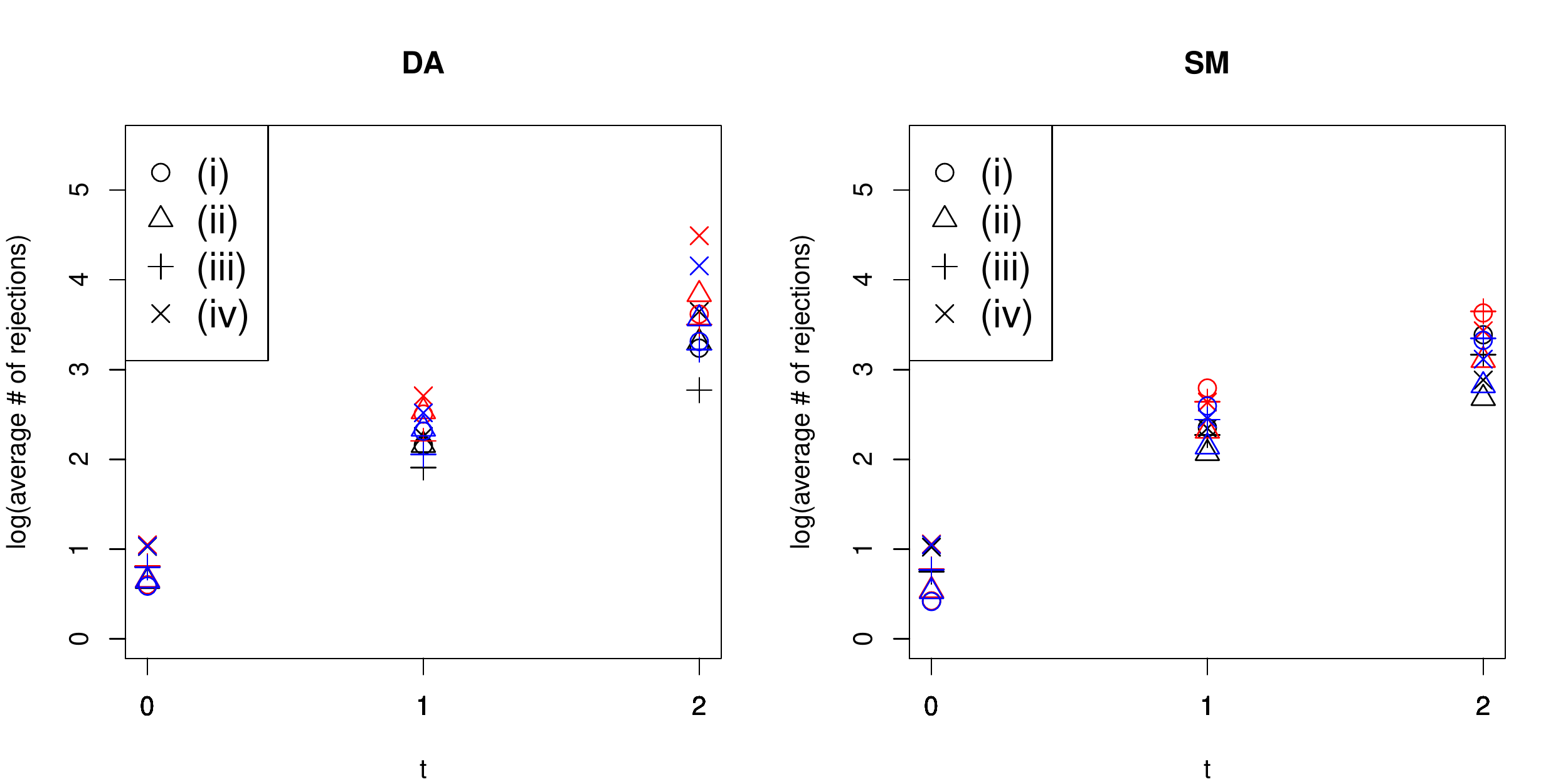}
\caption{Log average numbers of rejections per particle for DA and SM: $k=5$ (black), $k=10$ (red),  $k=10$ with summary (blue)}
\label{fig:sim1_1}
\end{figure}

\begin{figure}[H]
\centering
\includegraphics[width=\textwidth]{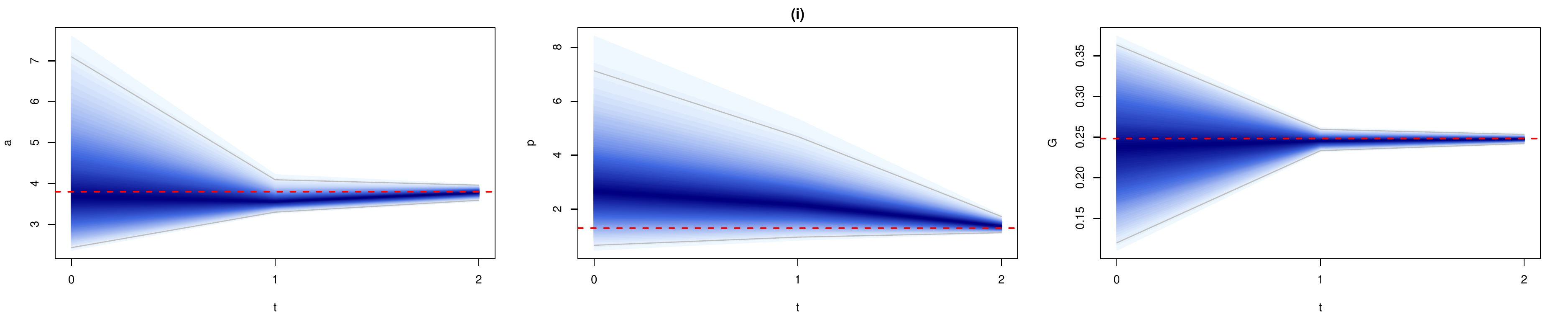}\\
\includegraphics[width=\textwidth]{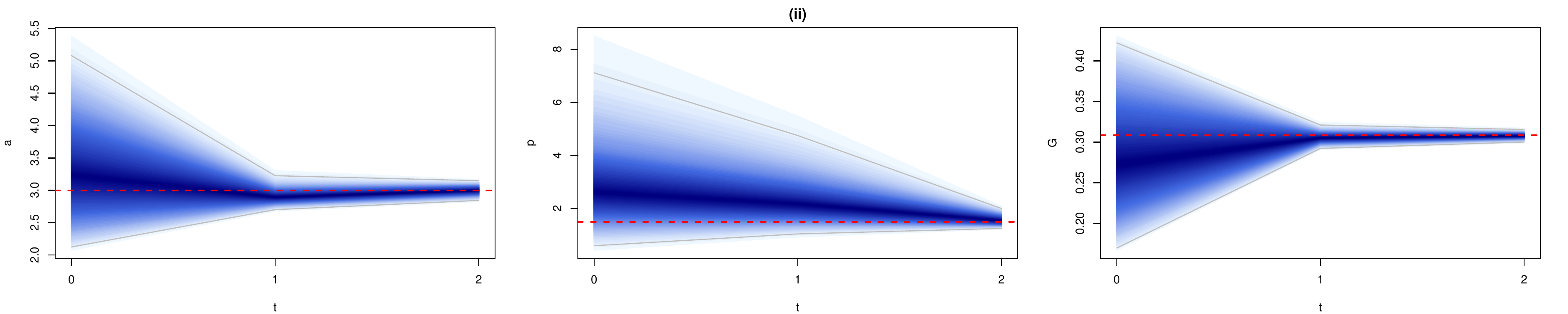}\\
\includegraphics[width=\textwidth]{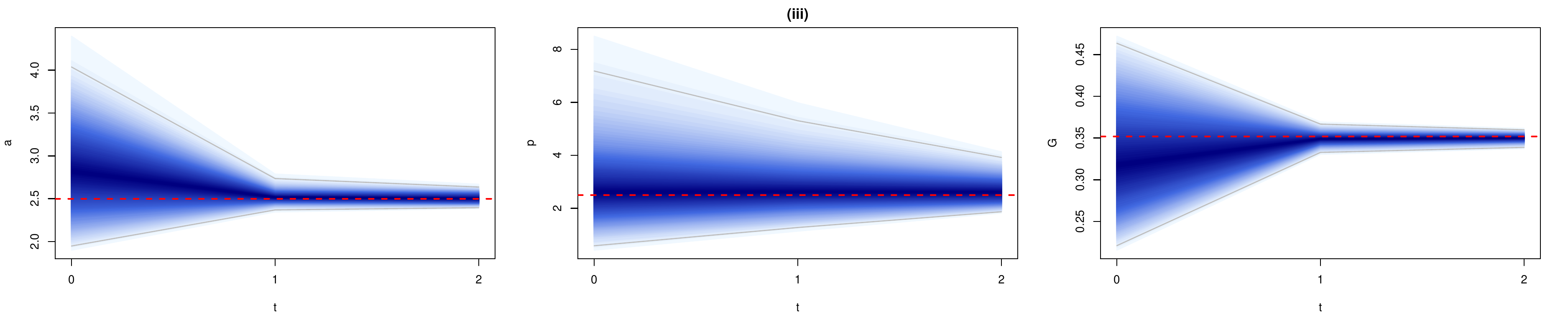}\\
\includegraphics[width=\textwidth]{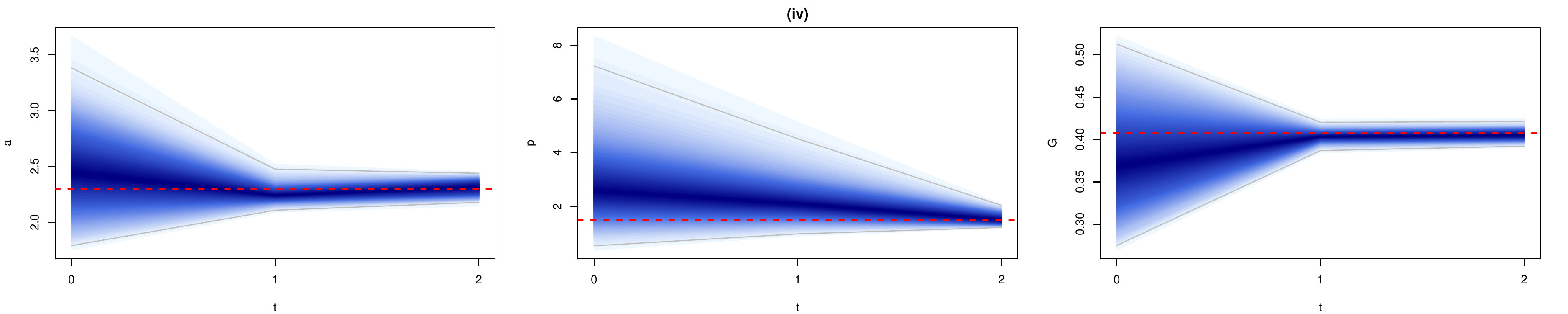}\\
\caption{Typical trajectories of Algorithm~\ref{alg} for DA ($k=5$) with the 2.5\% and 97.5\% quantiles (grey solid lines) and true parameter values (red dashed lines)}
\label{fig:sim1_2}
\end{figure}

\begin{figure}[H]
\centering
\includegraphics[width=\textwidth]{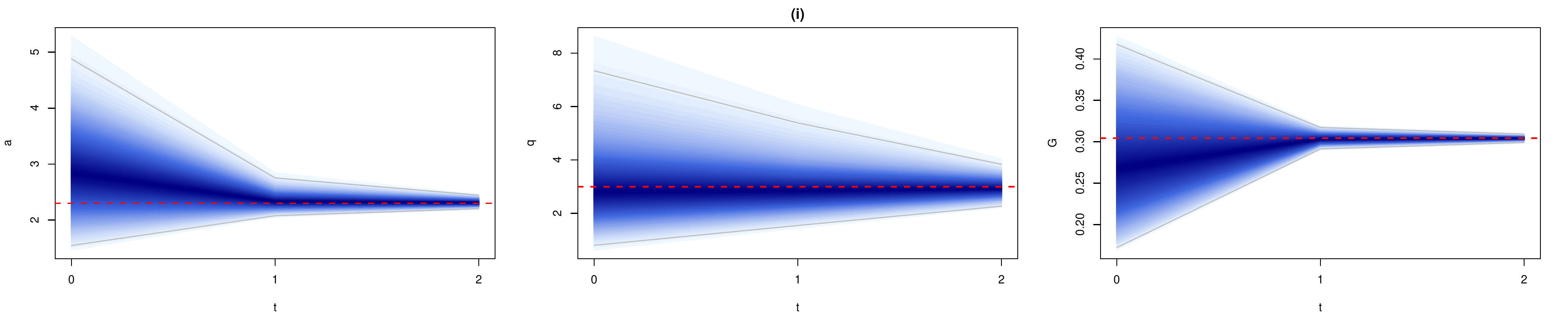}\\
\includegraphics[width=\textwidth]{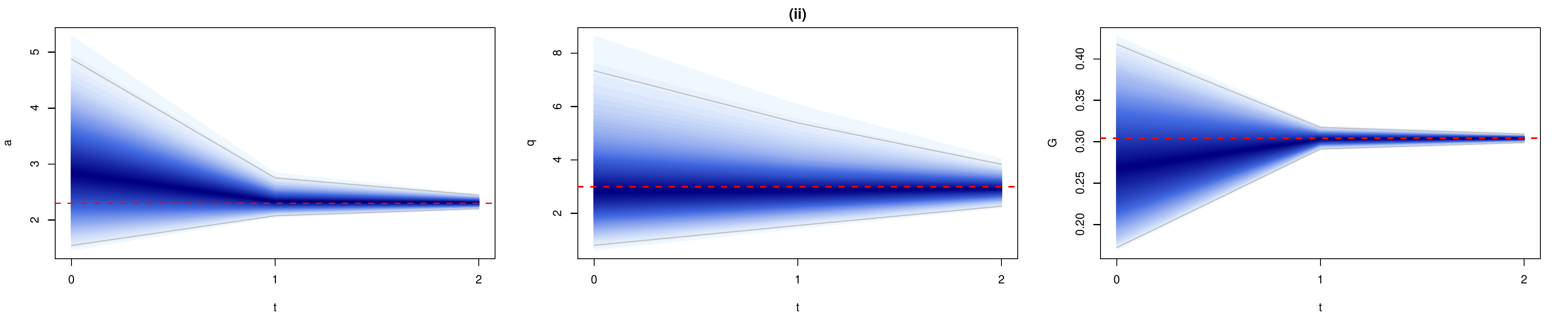}\\
\includegraphics[width=\textwidth]{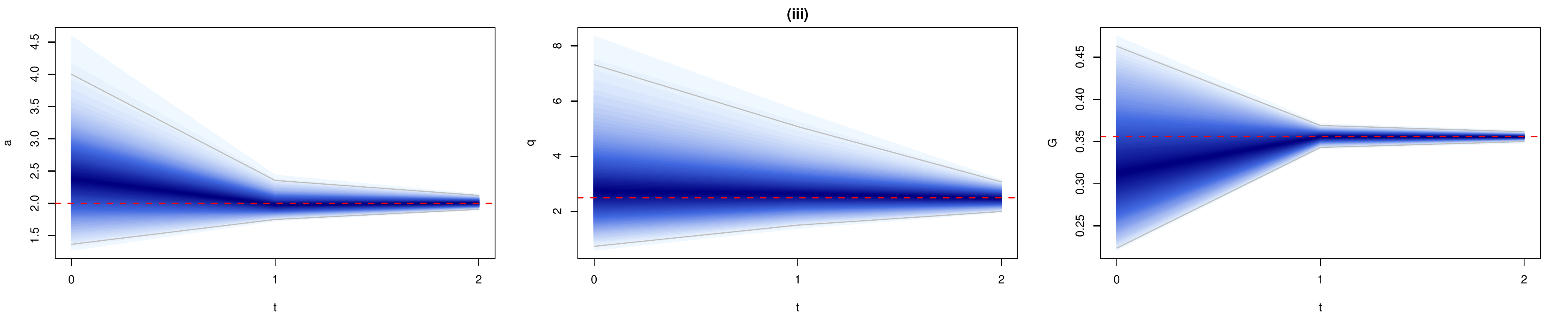}\\
\includegraphics[width=\textwidth]{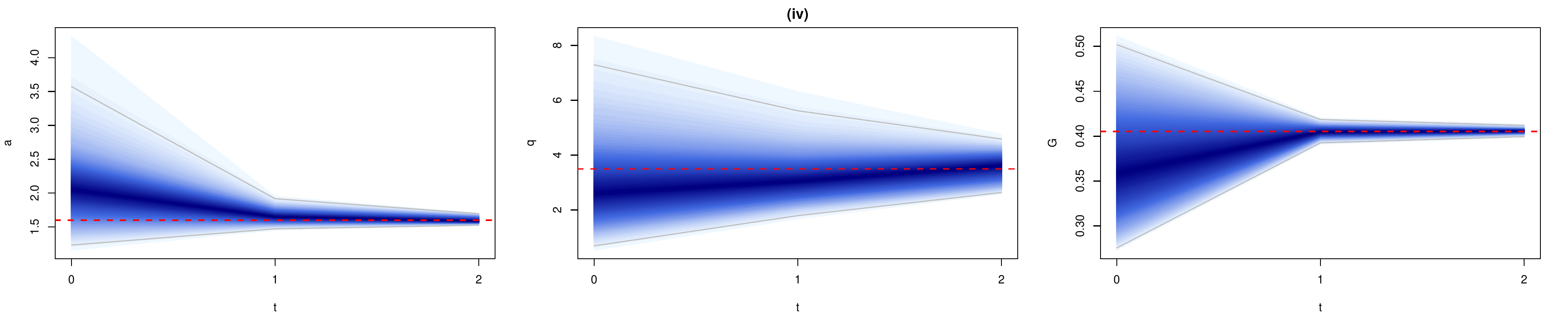}\\
\caption{Typical trajectories of Algorithm~\ref{alg} for SM ($k=5$) with the 2.5\% and 97.5\% quantiles (grey solid lines) and true parameter values (red dashed lines)}
\label{fig:sim1_3}
\end{figure}

Figure~\ref{fig:sim1_1} presents the log average numbers of rejections per particle for each step of Algorithm~\ref{alg} for DA and SM.
A large number of rejections implies longer computing time because the lines between \ref{alg:repeat} and \ref{alg:until} of Algorithm~\ref{alg} are repeated for an increased number of times.
The figure shows that the computing time of the algorithm increases as the number of groups increases in all cases.
Two nine-dimensional vectors are compared when $k=10$, leading to large numbers of rejections, but the computing time can be decreased by using the summary statistics.
In addition, the figure shows that the computing time may depend on the true Gini coefficient.
The average number of rejections tend to increase as the Gini coefficient increases in the case of DA while this tendency is less clear in the case of SM.
Figures~\ref{fig:sim1_2} and \ref{fig:sim1_3} present the typical trajectories of Algorithm~\ref{alg} for $k=5$ for DA and SM, respectively.
In the figures, the red horizontal dashed lines represent the true parameter values and the grey curves represent the 2.5\% and 97.5\% quantiles at each step. 
The figures show that the learning about the parameters and corresponding Gini coefficients occurs as the algorithm proceeds and the posterior distributions are concentrated around the true values under the target tolerance level.

Now, the performance of the three methods are compared.
Table~\ref{tab:sim1} presents the averages of the posterior means of the parameters and Gini coefficient and root mean squared errors (RMSE) for DA and SM over the 100 replicates.
Overall, ABC appears to work well. 
In the case of $k=5$, ABC resulted in the smallest RMSE for the Gini coefficients for both DA and SM. 
In the case of $k=10$, ABC and SOS produced the comparable result for DA and ABC and Dirichlet produced almost identical performance for SM in terms of the RMSE for the Gini coefficient.
As the available information increases from $k=5$ to $k=10$, the performance of ABC seems to improve slightly, but the degree of improvement is minuscule. 
On the other hand, we observe a clear improvement in the performance of SOS and Dirichlet as the number of income classes increases. 
The table also shows that the parameter and Gini estimates based on the Dirichlet likelihood in the case of $k=5$ are sensitive with respective to the prior specification for $\lambda$, while the sensitivity in the RMSE for the Gini coefficient vanishes in the case of $k=10$.

\renewcommand{\baselinestretch}{1.0}
\renewcommand{\arraystretch}{1.0}
\renewcommand{\tabcolsep}{1.5mm}
\begin{table}[H]
\centering
\caption{Result for DA and SM}\label{tab:sim1}
{\tiny
\begin{tabular}{cccclrrrrrrrrrrrr}\toprule
&&&&& \multicolumn{2}{c}{ABC}& \multicolumn{2}{c}{ABC (sum)} & \multicolumn{2}{c}{SOS} & \multicolumn{2}{c}{Dirichlet~1} & \multicolumn{2}{c}{Dirichlet~2} & \multicolumn{2}{c}{Dirichlet~3} \\\cmidrule(lr){6-7}\cmidrule(lr){8-9}\cmidrule(lr){10-11}\cmidrule(lr){12-13}\cmidrule(lr){14-15}\cmidrule(lr){16-17}
Model&$k$&Setting & Parameter & True & Mean & RMSE & Mean & RMSE & Mean & RMSE& Mean & RMSE& Mean & RMSE & Mean & RMSE \\\hline
DA           & 5  & (i)   & $a$ & 3.8    & 3.7898 & 0.0662 &        &        & 3.7689 & 0.1089 & 3.7598 & 0.1047 & 3.7319 & 0.1510 & 3.6908 & 0.1441 \\
             &    &       & $p$ & 1.3    & 1.3383 & 0.0861 &        &        & 1.3699 & 0.1584 & 1.4316 & 0.2655 & 1.7939 & 0.8557 & 1.9766 & 0.9693 \\
             &    &       & $G$ & 0.2482 & 0.2483 & 0.0023 &        &        & 0.2491 & 0.0035 & 0.2485 & 0.0052 & 0.2460 & 0.0067 & 0.2467 & 0.0039 \\
             &    & (ii)  & $a$ & 3.0    & 2.9973 & 0.0568 &        &        & 2.9743 & 0.0907 & 2.9967 & 0.0998 & 2.9652 & 0.0898 & 2.9690 & 0.1116 \\
             &    &       & $p$ & 1.5    & 1.5424 & 0.1192 &        &        & 1.6089 & 0.2771 & 1.5925 & 0.2461 & 1.9053 & 0.5965 & 2.0020 & 0.6270 \\
             &    &       & $G$ & 0.3087 & 0.3088 & 0.0034 &        &        & 0.3102 & 0.0050 & 0.3081 & 0.0062 & 0.3071 & 0.0066 & 0.3063 & 0.0073 \\
             &    & (iii) & $a$ & 2.5    & 2.4939 & 0.0506 &        &        & 2.4880 & 0.0696 & 2.5032 & 0.0510 & 2.5014 & 0.0541 & 2.4999 & 0.0499 \\
             &    &       & $p$ & 2.5    & 2.6983 & 0.3989 &        &        & 2.7611 & 0.6833 & 2.6459 & 0.3778 & 2.8400 & 0.4550 & 2.9155 & 0.5349 \\
             &    &       & $G$ & 0.3518 & 0.3523 & 0.0046 &        &        & 0.3532 & 0.0063 & 0.3512 & 0.0050 & 0.3508 & 0.0067 & 0.3514 & 0.0079 \\
             &    & (iv)  & $a$ & 2.3    & 2.2997 & 0.0506 &        &        & 2.2841 & 0.0644 & 2.2979 & 0.0636 & 2.2838 & 0.0583 & 2.2763 & 0.0639 \\
             &    &       & $p$ & 1.5    & 1.5478 & 0.1428 &        &        & 1.5889 & 0.2042 & 1.6119 & 0.3254 & 1.9052 & 0.7665 & 1.9121 & 0.5838 \\
             &    &       & $G$ & 0.4077 & 0.4081 & 0.0056 &        &        & 0.4098 & 0.0073 & 0.4070 & 0.0064 & 0.4055 & 0.0068 & 0.4064 & 0.0087 \\\cmidrule{2-17}
             & 10 & (i)   & $a$ & 3.8    & 3.7893 & 0.0704 & 3.7900 & 0.0726 & 3.7916 & 0.0640 & 3.8053 & 0.0709 & 3.7975 & 0.0703 & 3.7876 & 0.0711 \\
             &    &       & $p$ & 1.3    & 1.3374 & 0.0945 & 1.3396 & 0.0994 & 1.3185 & 0.0748 & 1.3023 & 0.0759 & 1.3296 & 0.0835 & 1.3664 & 0.1114 \\
             &    &       & $G$ & 0.2482 & 0.2483 & 0.0023 & 0.2483 & 0.0023 & 0.2485 & 0.0025 & 0.2481 & 0.0025 & 0.2479 & 0.0025 & 0.2478 & 0.0025 \\
             &    & (ii)  & $a$ & 3.0    & 2.9934 & 0.0584 & 3.0007 & 0.0637 & 2.9937 & 0.0498 & 3.0055 & 0.0621 & 3.0025 & 0.0622 & 2.9981 & 0.0621 \\
             &    &       & $p$ & 1.5    & 1.5494 & 0.1325 & 1.5361 & 0.1460 & 1.5247 & 0.0959 & 1.5043 & 0.1130 & 1.5280 & 0.1211 & 1.5608 & 0.1395 \\
             &    &       & $G$ & 0.3087 & 0.3090 & 0.0033 & 0.3088 & 0.0033 & 0.3092 & 0.0033 & 0.3084 & 0.0037 & 0.3083 & 0.0037 & 0.3082 & 0.0037 \\
             &    & (iii) & $a$ & 2.5    & 2.4945 & 0.0488 & 2.4939 & 0.0510 & 2.4960 & 0.0395 & 2.5068 & 0.0590 & 2.5052 & 0.0583 & 2.5035 & 0.0576 \\
             &    &       & $p$ & 2.5    & 2.6868 & 0.4077 & 2.7127 & 0.4477 & 2.5705 & 0.2433 & 2.5309 & 0.3753 & 2.5866 & 0.3878 & 2.6526 & 0.4094 \\
             &    &       & $G$ & 0.3518 & 0.3523 & 0.0043 & 0.3524 & 0.0044 & 0.3524 & 0.0040 & 0.3514 & 0.0054 & 0.3513 & 0.0054 & 0.3511 & 0.0055 \\
             &    & (iv)  & $a$ & 2.3    & 2.2962 & 0.0477 & 2.2978 & 0.0521 & 2.2960 & 0.0382 & 2.3074 & 0.0591 & 2.3059 & 0.0589 & 2.3043 & 0.0589 \\
             &    &       & $p$ & 1.5    & 1.5560 & 0.1471 & 1.5643 & 0.1660 & 1.5229 & 0.0954 & 1.5034 & 0.1528 & 1.5206 & 0.1579 & 1.5411 & 0.1664 \\
             &    &       & $G$ & 0.4077 & 0.4084 & 0.0051 & 0.4083 & 0.0054 & 0.4084 & 0.0047 & 0.4071 & 0.0065 & 0.4070 & 0.0065 & 0.4068 & 0.0066 \\\hline
SM           & 5  & (i)   & $a$ & 3.5    & 3.4741 & 0.0587 &        &        & 3.4684 & 0.0993 & 3.4596 & 0.1355 & 3.3927 & 0.2073 & 3.3509 & 0.1995 \\
             &    &       & $q$ & 1.5    & 1.5526 & 0.0906 &        &        & 1.6054 & 0.2149 & 1.6111 & 0.2454 & 1.8799 & 0.6104 & 2.0986 & 0.8541 \\
             &    &       & $G$ & 0.2429 & 0.2430 & 0.0018 &        &        & 0.2424 & 0.0038 & 0.2428 & 0.0038 & 0.2422 & 0.0032 & 0.2410 & 0.0067 \\
             &    & (ii)  & $a$ & 2.3    & 2.2913 & 0.0351 &        &        & 2.2939 & 0.0537 & 2.2939 & 0.0351 & 2.3127 & 0.0756 & 2.3116 & 0.0512 \\
             &    &       & $q$ & 3.0    & 3.1518 & 0.2884 &        &        & 3.3001 & 0.6636 & 3.1403 & 0.3077 & 3.2302 & 0.3546 & 3.3163 & 0.3836 \\
             &    &       & $G$ & 0.3041 & 0.3042 & 0.0021 &        &        & 0.3042 & 0.0041 & 0.3042 & 0.0033 & 0.3032 & 0.0035 & 0.3032 & 0.0034 \\
             &    & (iii) & $a$ & 2.0    & 1.9916 & 0.0337 &        &        & 1.9893 & 0.0505 & 1.9904 & 0.0325 & 1.9938 & 0.0737 & 1.9899 & 0.0346 \\
             &    &       & $q$ & 2.5    & 2.6054 & 0.2182 &        &        & 2.7626 & 0.5330 & 2.6295 & 0.2697 & 2.7826 & 0.6686 & 2.8145 & 0.4163 \\
             &    &       & $G$ & 0.3567 & 0.3557 & 0.0026 &        &        & 0.3552 & 0.0056 & 0.3553 & 0.0033 & 0.3555 & 0.0097 & 0.3549 & 0.0042 \\
             &    & (iv)  & $a$ & 1.6    & 1.5934 & 0.0271 &        &        & 1.6019 & 0.0350 & 1.6061 & 0.0408 & 1.6101 & 0.0414 & 1.6272 & 0.0731 \\
             &    &       & $q$ & 3.5    & 3.7155 & 0.4137 &        &        & 3.7560 & 0.7196 & 3.5763 & 0.3779 & 3.6418 & 0.3381 & 3.6086 & 0.3967 \\
             &    &       & $G$ & 0.4052 & 0.4051 & 0.0028 &        &        & 0.4060 & 0.0059 & 0.4048 & 0.0032 & 0.4046 & 0.0037 & 0.4038 & 0.0044 \\\cmidrule{2-17}
             & 10 & (i)   & $a$ & 3.5    & 3.4806 & 0.0592 & 3.4838 & 0.0603 & 3.4860 & 0.0659 & 3.4931 & 0.0519 & 3.4825 & 0.0532 & 3.4707 & 0.0574 \\
             &    &       & $q$ & 1.5    & 1.5397 & 0.0856 & 1.5371 & 0.0863 & 1.5389 & 0.1076 & 1.5167 & 0.0711 & 1.5385 & 0.0802 & 1.5680 & 0.1002 \\
             &    &       & $G$ & 0.2429 & 0.2430 & 0.0019 & 0.2430 & 0.0019 & 0.2427 & 0.0024 & 0.2428 & 0.0020 & 0.2428 & 0.0019 & 0.2428 & 0.0019 \\
             &    & (ii)  & $a$ & 2.3    & 2.2941 & 0.0356 & 2.2941 & 0.0369 & 2.2947 & 0.0408 & 2.2961 & 0.0311 & 2.2952 & 0.0313 & 2.2944 & 0.0309 \\
             &    &       & $q$ & 3.0    & 3.1075 & 0.2617 & 3.1133 & 0.2742 & 3.1434 & 0.3879 & 3.0558 & 0.2163 & 3.0984 & 0.2384 & 3.1396 & 0.2583 \\
             &    &       & $G$ & 0.3041 & 0.3042 & 0.0021 & 0.3042 & 0.0021 & 0.3040 & 0.0028 & 0.3041 & 0.0021 & 0.3040 & 0.0021 & 0.3039 & 0.0021 \\
             &    & (iii) & $a$ & 2.0    & 1.9951 & 0.0344 & 1.9958 & 0.0365 & 1.9941 & 0.0362 & 1.9973 & 0.0288 & 1.9956 & 0.0293 & 1.9939 & 0.0293 \\
             &    &       & $q$ & 2.5    & 2.5700 & 0.1996 & 2.5733 & 0.2127 & 2.6047 & 0.2784 & 2.5339 & 0.1623 & 2.5592 & 0.1743 & 2.5859 & 0.1869 \\
             &    &       & $G$ & 0.3567 & 0.3557 & 0.0025 & 0.3557 & 0.0025 & 0.3554 & 0.0036 & 0.3556 & 0.0025 & 0.3555 & 0.0025 & 0.3555 & 0.0026 \\
             &    & (iv)  & $a$ & 1.6    & 1.5968 & 0.0281 & 1.5950 & 0.0306 & 1.5985 & 0.0275 & 1.5985 & 0.0245 & 1.5980 & 0.0248 & 1.5997 & 0.0227 \\
             &    &       & $q$ & 3.5    & 3.6390 & 0.3775 & 3.6758 & 0.4182 & 3.6640 & 0.4959 & 3.5663 & 0.3118 & 3.6022 & 0.3362 & 3.6034 & 0.3074 \\
             &    &       & $G$ & 0.4052 & 0.4051 & 0.0027 & 0.4052 & 0.0027 & 0.4050 & 0.0039 & 0.4050 & 0.0027 & 0.4049 & 0.0027 & 0.4048 & 0.0027 \\\bottomrule
\end{tabular}
}

\begin{minipage}{350pt}
{\tiny
ABC (sum) denotes the ABC method using the summary statistics when $k=10$.
Dirichlet 1, 2, and 3 denote the reuslts under the Dirichlet likelihood with $\G(1,0.1)$, $\G(1,0.5)$, and $\G(1,1.0)$ priors for $\lambda$, respectively.
}
\end{minipage}
\end{table}

\subsection{Simulated Data 2}\label{sec:sim2}
To study the potential of the proposed approach, the more flexible alternatives to the Dagum and Singh-Maddala distributions, namely the generalised beta distribution of the second kind (GB2), denoted by $\GB2(a,b,p,q)=\GB(a,b,1,p,q)$, and the five-parameter GB distribution, are additionally considered.
For GB2, the data are generated from $\GB2(a,1,p,q)$ based on the following four settings:
(i) $(a,p,q,G)=(2.5,2.3,1.7,0.2572)$,
(ii) $(a,p,q,G)=(2.1,1.8,2.0,0.3037)$,
(iii) $(a,p,q,G)=(1.8,3.0,1.5,0.3536)$,
(iv) $(a,p,q,G)=(1.5,2.5,1.8,0.4064)$.
For GB, the following five settings covering various values of the parameters and Gini coefficient are considered:
(i) $(a,c,p,q,G)=(2.0,0.95,3.0,2.0,0.2456)$,
(ii) $(a,c,p,q,G)=(1.2,0.4,1.7,2.5,0.3062)$,
(iii) $(a,c,p,q,G)=(1.5,0.9,1.7,1.7,0.3589)$,
(iv) $(a,c,p,q,G)=(1.2,0.1,1.3,3.5,0.3397)$,
(v) $(a,c,p,q,G)=(1.5,0.99,1.2,3.0,0.4105)$.
The data are replicated 50 times.
We implement only ABC and SOS, because the analytical form of the Lorenz curves for GB and GB2 are unknown. 
In addition to the prior distributions for $a$, $p$, $q$ specified in Section~\ref{sec:sim1}, $c\sim\U(0,1)$ is assumed. 
The MCMC algorithm for SOS is run for $70000$ iterations including the $10000$ initial burn-in period and every 20th draw is retained for posterior inference.
For ABC, the same setting for Algorith~\ref{alg} as in Section~\ref{sec:sim1} is used.

\begin{figure}[H]
\centering
\includegraphics[width=\textwidth]{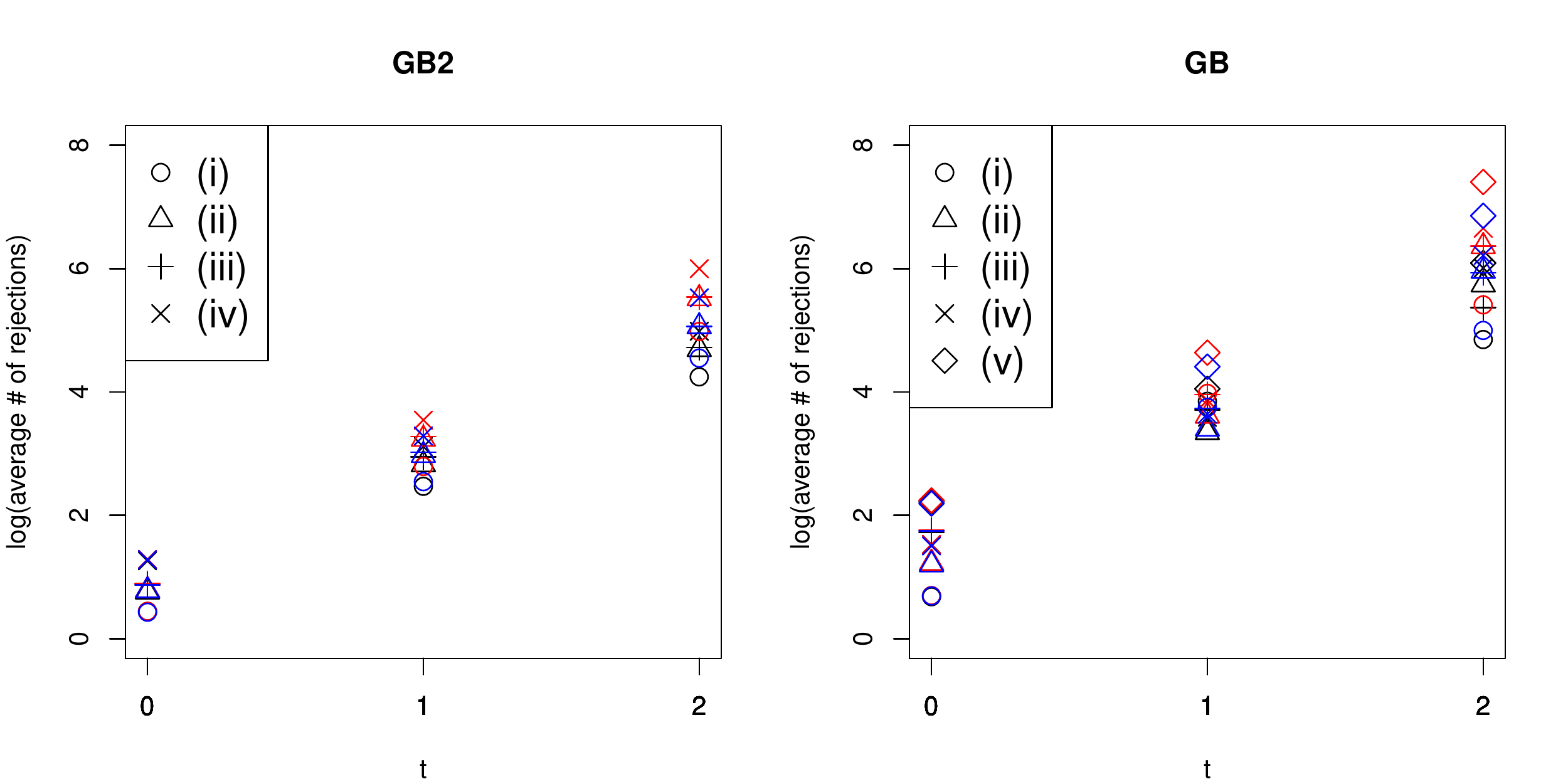}
\caption{Log average number of rejections per particle for GB2 and GB: $k=5$ (black), $k=10$ (red),  $k=10$ with summary (blue)}
\label{fig:sim2_1}
\end{figure}

Figure~\ref{fig:sim2_1} presents the log average numbers of rejections per particle for each step of Algorithm~\ref{alg} for GB and GB2.
The figure shows that the overall numbers of rejections are larger for GB and GB2 than for DA and SM leading to the increased computing time, since it is required to estimate more parameters for GB and GB2.
The average number of rejections tends to increase when the true Gini coefficient increases.
In addition, using the summary statistics in the case of $k=10$ results in the shorter computing time. 

\begin{figure}[H]
\centering
\includegraphics[width=\textwidth]{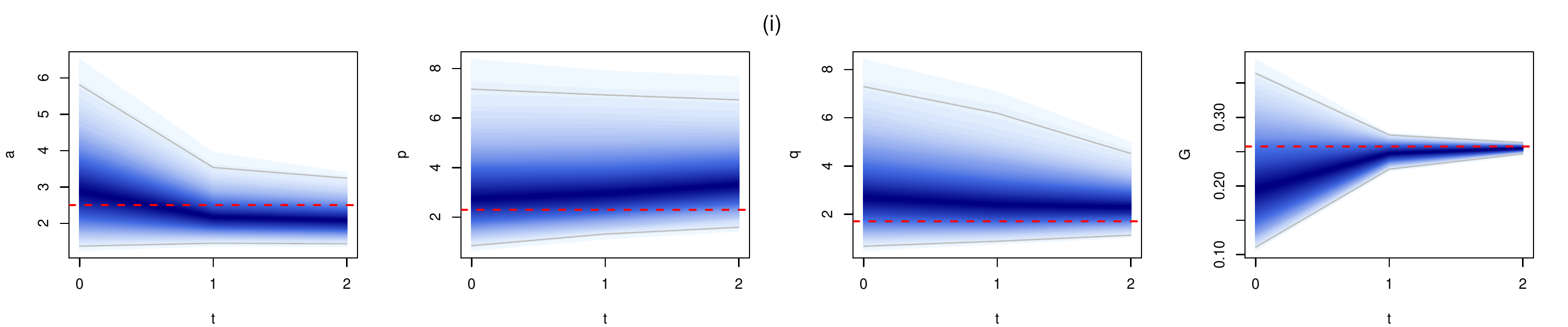}\\
\includegraphics[width=\textwidth]{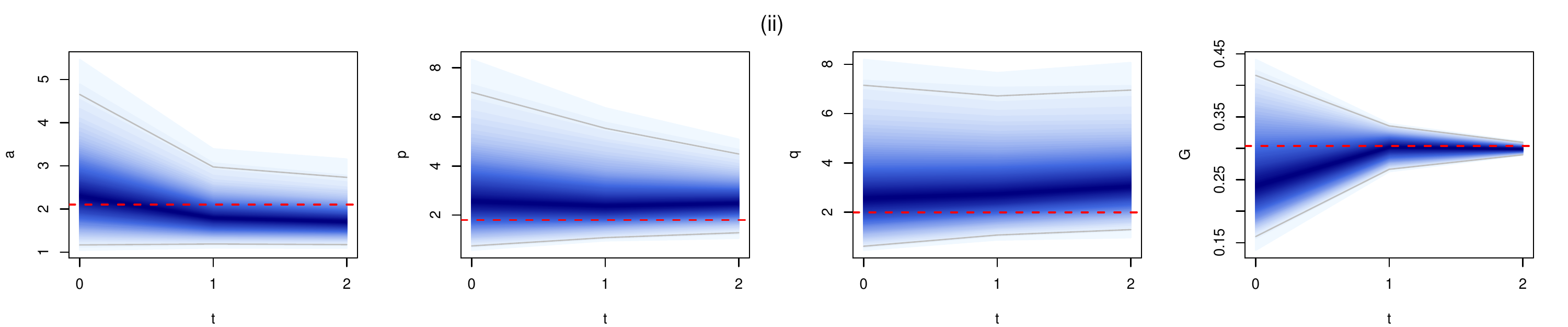}\\
\includegraphics[width=\textwidth]{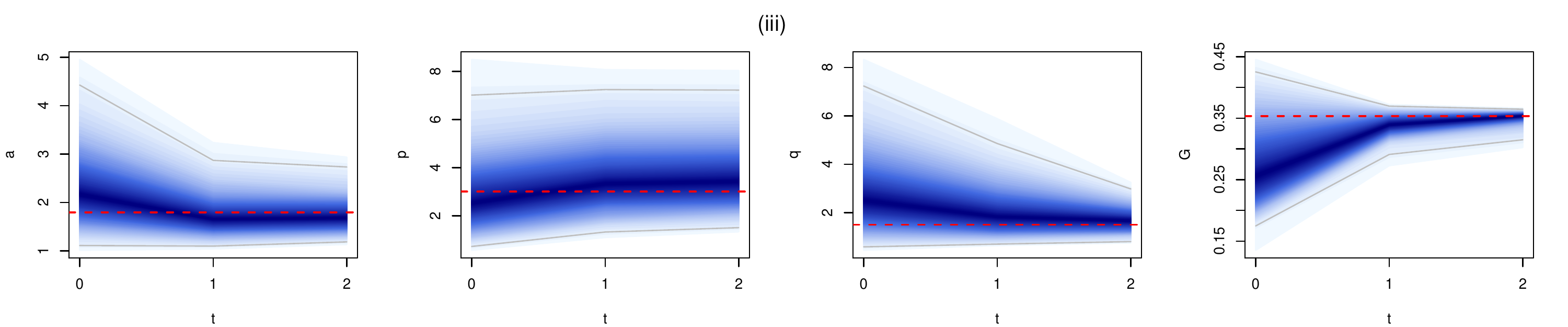}\\
\includegraphics[width=\textwidth]{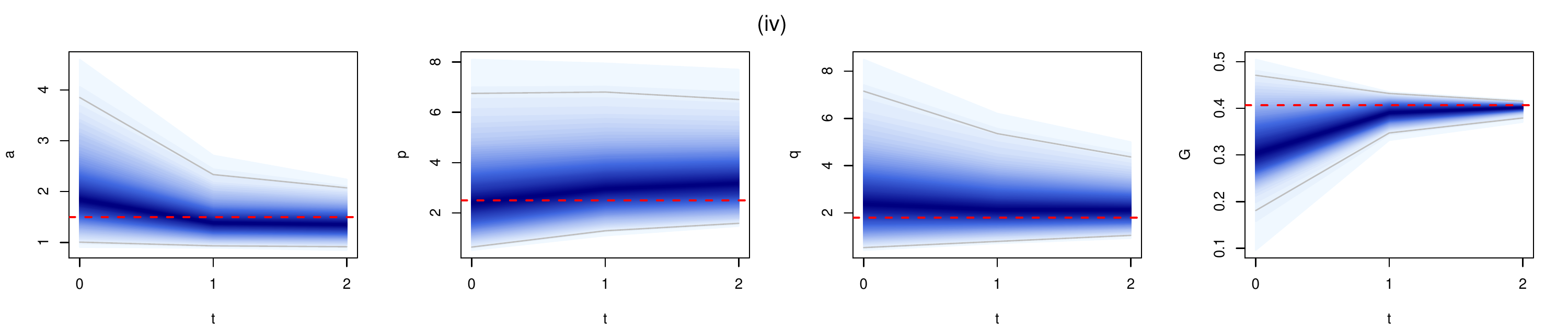}\\
\caption{Typical trajectories of Algorithm~\ref{alg} for GB2 ($k=5$) with the 2.5\% and 97.5\% quantiles (grey solid lines) and true parameter values (grey dashed lines)}
\label{fig:sim2_2}
\end{figure}

\begin{figure}[H]
\centering
\includegraphics[width=\textwidth]{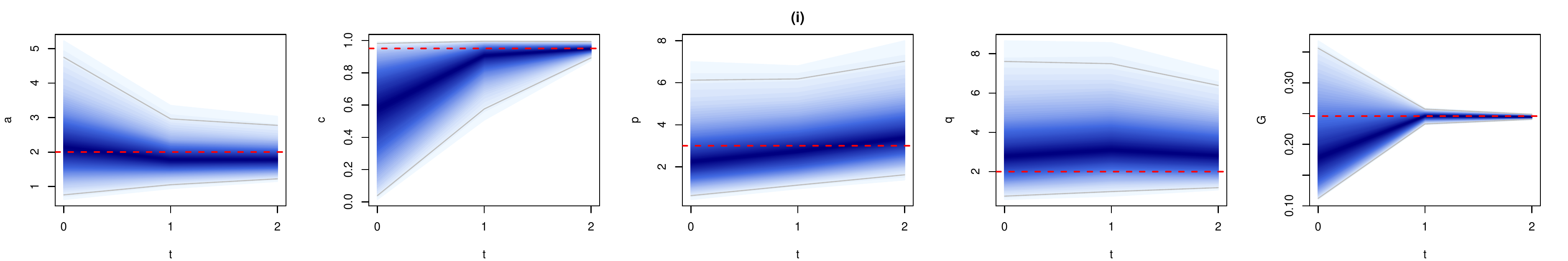}\\
\includegraphics[width=\textwidth]{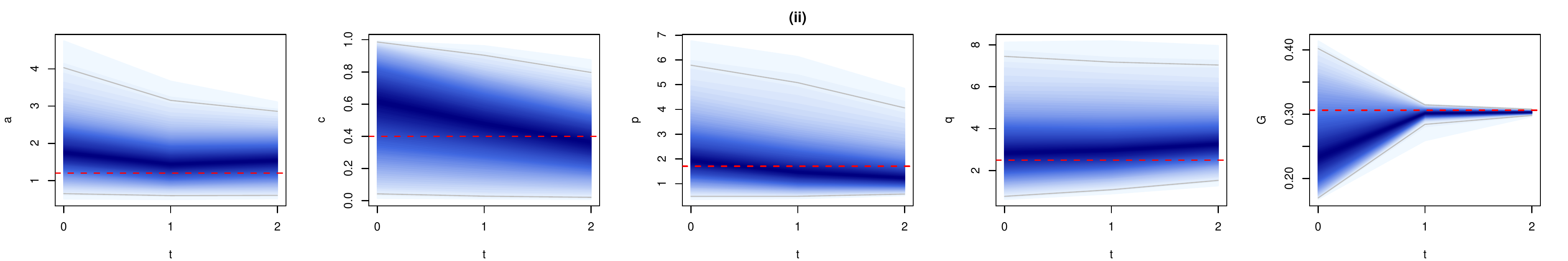}\\
\includegraphics[width=\textwidth]{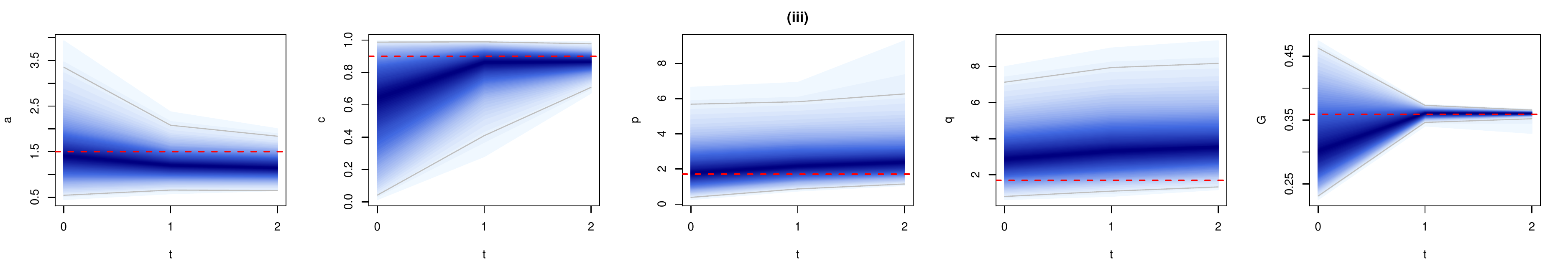}\\
\includegraphics[width=\textwidth]{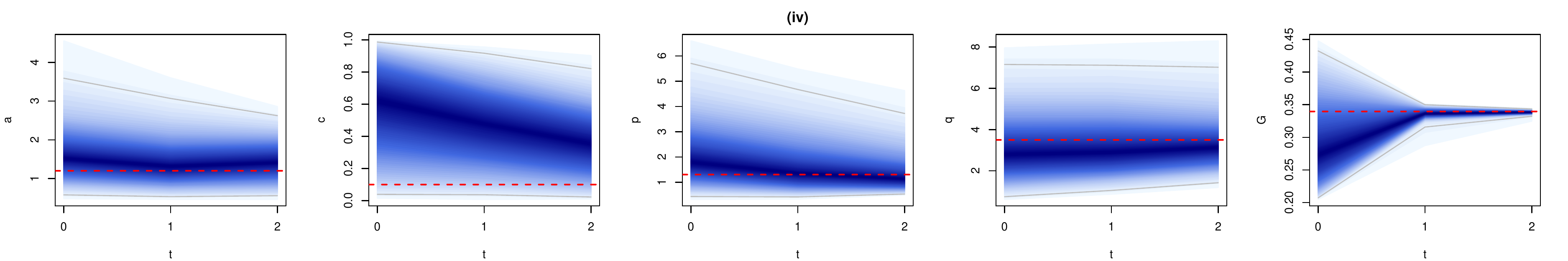}\\
\includegraphics[width=\textwidth]{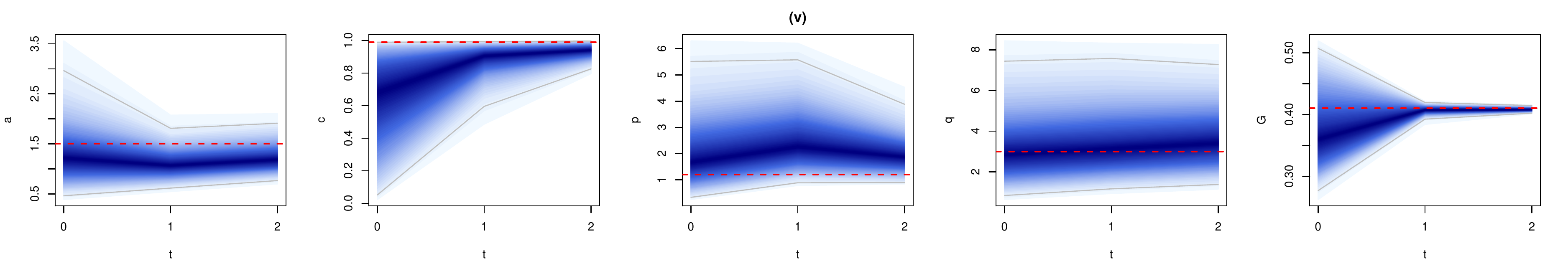}\\
\caption{Typical trajectories of Algorithm~\ref{alg} for GB ($k=5$) with the 2.5\% and 97.5\% quantiles (grey solid lines) and true parameter values (red dashed lines)}
\label{fig:sim2_3}
\end{figure}

Figures~\ref{fig:sim2_2} and \ref{fig:sim2_3} present the typical trajectories of Algorithm~\ref{alg} for $k=5$ for GB2 and GB, respectively.
The red horizontal dashed lines represent the true parameter values and the grey curves represent the 2.5\% and 97.5\% quantiles. 
In contrast to the cases of DA and SM, the figure shows that not all parameters are simultaneously identified from the data, because the information contained in the data is limieted. 
For example, in Setting~(i) for GB2, Figure~\ref{fig:sim2_2} shows that the learning about $a$ and $q$ occurs and the posterior distributions concentrate as the algorithm proceeds, but little learning about $p$ occurs. 
Similarly, for GB, Figure~\ref{fig:sim2_3} shows we can only learn about $a$ and $c$ in Setting~(i) and about $a$ and $p$ in Setting~(iv). 
Which parameters we can learn seems to depend on the simulation setting. 
Nonetheless, the figures also show that in all cases the learning about the Gini coefficient does occur as the algorithm proceeds and the posterior distributions under the target tolerance are concentrated around the true values.

Table~\ref{tab:sim2} presents the averages of the posterior mean and RMSE for the parameters and the Gini coefficient for GB2 and GB under the two methods.
In all cases, ABC produced the smaller RMSE for the Gini coefficient than SOS. 
For both methods, the performance improves as the number of groups increases, but the degree of improvement is minuscule for ABC compared to SOS, as in the cases of DA and SM.
The large RMSEs for the parameters in the table corresponds to the cases where the parameters are not well identified from the data as shown by Figures~\ref{fig:sim2_2} and \ref{fig:sim2_3}. 
The large RMSE for SOS could be also attributed to the poor mixing and convergence failure of the MCMC algorithm, as the convergence of MCMC in the context of grouped data is typically difficult to ensure (Kakamu,~2016).

\renewcommand{\baselinestretch}{1.0}
\renewcommand{\arraystretch}{1.0}
\renewcommand{\tabcolsep}{1mm}
\begin{table}[H]
\centering
\caption{Result for GB2 and GB distributions}\label{tab:sim2}
{\scriptsize
\begin{tabular}{lcclrrrrrrrrrr}\toprule
&&&& \multicolumn{4}{c}{$k=5$}&\multicolumn{6}{c}{$k=10$}\\\cmidrule(lr){5-8}\cmidrule(lr){9-14}
&&&& \multicolumn{2}{c}{ABC} & \multicolumn{2}{c}{SOS} & \multicolumn{2}{c}{ABC}  & \multicolumn{2}{c}{ABC(sum)} & \multicolumn{2}{c}{SOS} \\\cmidrule(lr){5-6}\cmidrule(lr){7-8}\cmidrule(lr){9-10}\cmidrule(lr){11-12}\cmidrule(lr){13-14}
Model      & Setting & Parameter & True & Mean & RMSE & Mean & RMSE& Mean & RMSE & Mean & RMSE & Mean & RMSE\\\hline
GB2        & (i)   & $a$ & 2.5    & 2.1391 & 0.3630 & 2.2944 & 0.4645 & 2.1610 & 0.3473 & 2.1572 & 0.3497 & 2.3658 & 0.4295 \\
           &       & $p$ & 2.3    & 3.3641 & 1.0741 & 2.1571 & 0.6206 & 3.2156 & 0.9449 & 3.2277 & 0.9545 & 2.0440 & 0.6048 \\
           &       & $q$ & 1.7    & 2.5028 & 0.8085 & 3.2220 & 1.8285 & 2.4675 & 0.7792 & 2.4978 & 0.8067 & 2.9374 & 1.6008 \\
           &       & $G$ & 0.2572 & 0.2588 & 0.0029 & 0.2482 & 0.0099 & 0.2584 & 0.0026 & 0.2582 & 0.0026 & 0.2493 & 0.0085 \\
           & (ii)  & $a$ & 2.1    & 1.8001 & 0.3066 & 1.9395 & 0.3392 & 1.8325 & 0.2848 & 1.8301 & 0.2843 & 1.9047 & 0.3653 \\
           &       & $p$ & 1.8    & 2.4669 & 0.6866 & 2.5965 & 1.0547 & 2.3497 & 0.5920 & 2.3524 & 0.5904 & 2.6887 & 1.1674 \\
           &       & $q$ & 2.0    & 3.1481 & 1.1573 & 2.3008 & 0.6511 & 3.0366 & 1.0690 & 3.0634 & 1.0853 & 2.3170 & 0.6595 \\
           &       & $G$ & 0.3037 & 0.3027 & 0.0031 & 0.3093 & 0.0084 & 0.3022 & 0.0033 & 0.3021 & 0.0033 & 0.3096 & 0.0077 \\
           & (iii) & $a$ & 1.8    & 1.7017 & 0.1076 & 1.7334 & 0.2945 & 1.6945 & 0.1275 & 1.6907 & 0.1301 & 1.7386 & 0.2506 \\
           &       & $p$ & 3.0    & 3.7787 & 0.8002 & 1.7688 & 1.3589 & 3.6963 & 0.7577 & 3.7207 & 0.7754 & 1.6814 & 1.3617 \\
           &       & $q$ & 1.5    & 1.8040 & 0.3212 & 3.8621 & 2.7102 & 1.8054 & 0.3337 & 1.8267 & 0.3550 & 3.8035 & 2.6010 \\
           &       & $G$ & 0.3536 & 0.3721 & 0.0192 & 0.3284 & 0.0259 & 0.3715 & 0.0187 & 0.3713 & 0.0185 & 0.3280 & 0.0259 \\
           & (iv)  & $a$ & 1.5    & 1.3756 & 0.1311 & 1.4624 & 0.2812 & 1.3900 & 0.1374 & 1.3875 & 0.1357 & 1.3864 & 0.2359 \\
           &       & $p$ & 2.5    & 3.2409 & 0.7699 & 2.0663 & 0.7222 & 3.1050 & 0.6968 & 3.1176 & 0.6985 & 2.2012 & 0.6111 \\
           &       & $q$ & 1.8    & 2.3654 & 0.5745 & 3.2108 & 1.8263 & 2.3084 & 0.5468 & 2.3326 & 0.5635 & 3.3054 & 1.8158 \\
           &       & $G$ & 0.4064 & 0.4138 & 0.0085 & 0.3863 & 0.0221 & 0.4134 & 0.0081 & 0.4132 & 0.0080 & 0.3862 & 0.0208 \\\hline
GB         & (i)   & $a$ & 2.0    & 1.8520 & 0.1494 & 2.1262 & 0.2347 & 1.8109 & 0.1910 & 1.8212 & 0.1833 & 2.0275 & 0.2315 \\
           &       & $c$ & 0.95   & 0.9538 & 0.0060 & 0.9596 & 0.0139 & 0.9490 & 0.0048 & 0.9489 & 0.0051 & 0.9608 & 0.0157 \\
           &       & $p$ & 3.0    & 3.6033 & 0.6149 & 2.8358 & 0.6684 & 3.6951 & 0.7078 & 3.6701 & 0.6980 & 3.0564 & 0.5842 \\
           &       & $q$ & 2.0    & 3.0596 & 1.0637 & 2.0725 & 0.5179 & 3.1261 & 1.1391 & 3.1288 & 1.1399 & 2.3028 & 0.5987 \\
           &       & $G$ & 0.2456 & 0.2462 & 0.0017 & 0.2465 & 0.0060 & 0.2461 & 0.0016 & 0.2461 & 0.0016 & 0.2477 & 0.0039 \\
           & (ii)  & $a$ & 1.2    & 1.4097 & 0.2276 & 1.4868 & 0.3756 & 1.3348 & 0.1646 & 1.3531 & 0.1756 & 1.4527 & 0.4268 \\
           &       & $c$ & 0.4    & 0.3698 & 0.0338 & 0.3990 & 0.1252 & 0.3418 & 0.0616 & 0.3466 & 0.0568 & 0.3912 & 0.1295 \\
           &       & $p$ & 1.7    & 1.7556 & 0.1512 & 1.3711 & 0.4350 & 1.8399 & 0.2225 & 1.8189 & 0.1915 & 1.4707 & 0.4757 \\
           &       & $q$ & 2.5    & 3.2228 & 0.7494 & 3.2719 & 1.0712 & 3.0909 & 0.6281 & 3.0885 & 0.6222 & 3.1895 & 1.0178 \\
           &       & $G$ & 0.3062 & 0.3049 & 0.0022 & 0.3080 & 0.0037 & 0.3050 & 0.0021 & 0.3049 & 0.0022 & 0.3067 & 0.0027 \\
           & (iii) & $a$ & 1.5    & 1.1748 & 0.3267 & 1.4081 & 0.1884 & 1.1168 & 0.3842 & 1.1197 & 0.3813 & 1.2915 & 0.3140 \\
           &       & $c$ & 0.9    & 0.8484 & 0.0530 & 0.9067 & 0.0351 & 0.8250 & 0.0762 & 0.8240 & 0.0772 & 0.8929 & 0.0359 \\
           &       & $p$ & 1.7    & 2.6511 & 0.9607 & 1.8787 & 0.3954 & 2.8834 & 1.1949 & 2.8755 & 1.1871 & 2.2437 & 0.8159 \\
           &       & $q$ & 1.7    & 3.7431 & 2.0464 & 2.6466 & 1.1092 & 3.5490 & 1.8728 & 3.5753 & 1.8998 & 3.0134 & 1.5576 \\
           &       & $G$ & 0.3589 & 0.3590 & 0.0022 & 0.3648 & 0.0090 & 0.3584 & 0.0022 & 0.3583 & 0.0024 & 0.3630 & 0.0065 \\
           & (iv)  & $a$ & 1.2    & 1.4435 & 0.2658 & 1.3013 & 0.2700 & 1.4049 & 0.2448 & 1.4117 & 0.2532 & 1.3013 & 0.2700\\
           &       & $c$ & 0.1    & 0.3384 & 0.2389 & 0.2772 & 0.2016 & 0.3022 & 0.2027 & 0.3092 & 0.2095 & 0.2772 & 0.2016\\
           &       & $p$ & 1.3    & 1.4112 & 0.1812 & 1.3132 & 0.3282 & 1.4078 & 0.2036 & 1.4194 & 0.2139 & 1.3132 & 0.3282\\
           &       & $q$ & 3.5    & 3.1348 & 0.4202 & 3.2402 & 0.7659 & 3.1263 & 0.4480 & 3.1159 & 0.4599 & 3.2402 & 0.7659\\
           &       & $G$ & 0.3397 & 0.3379 & 0.0028 & 0.3390 & 0.0052 & 0.3384 & 0.0024 & 0.3383 & 0.0024 & 0.3390 & 0.0052\\
           & (v)   & $a$ & 1.5    & 1.1889 & 0.3176 & 1.2606 & 0.2998 & 1.2986 & 0.2260 & 1.2841 & 0.2367 & 1.3753 & 0.2845 \\
           &       & $c$ & 0.99   & 0.9123 & 0.0792 & 0.8617 & 0.1405 & 0.9474 & 0.0462 & 0.9452 & 0.0483 & 0.9206 & 0.0810 \\
           &       & $p$ & 1.2    & 2.0412 & 0.8583 & 1.6687 & 0.5761 & 1.7250 & 0.5649 & 1.7807 & 0.6166 & 1.4961 & 0.5021 \\
           &       & $q$ & 3.0    & 3.7646 & 0.7763 & 3.3483 & 0.8098 & 3.6695 & 0.7539 & 3.6939 & 0.7594 & 3.3561 & 1.1940 \\
           &       & $G$ & 0.4105 & 0.4096 & 0.0029 & 0.3968 & 0.0153 & 0.4105 & 0.0028 & 0.4105 & 0.0028 & 0.4043 & 0.0075 \\\bottomrule
\end{tabular}

\begin{minipage}{200pt}
{\tiny
ABC(sum) denotes the ABC method using the summary statistics.
}
\end{minipage}
}
\end{table}

\subsection{Real Data: Family Income and Expenditure Survey in Japan}
The proposed method is now applied to estimate the Gini coefficient of the data from the Family Income and Expenditure Survey (FIES) in 2012 prepared by Ministry of Internal Affairs and Communications of Japan.
The FIES data are based on $n=10000$ households and are available in the forms of quintile and decile data.
For the hypothetical income distributions, DA, SM, GB2, and GB are fitted. 
The same prior distributions and algorithm settings as in the simulation studies are used.

\begin{figure}[H]
\centering
\includegraphics[width=\textwidth]{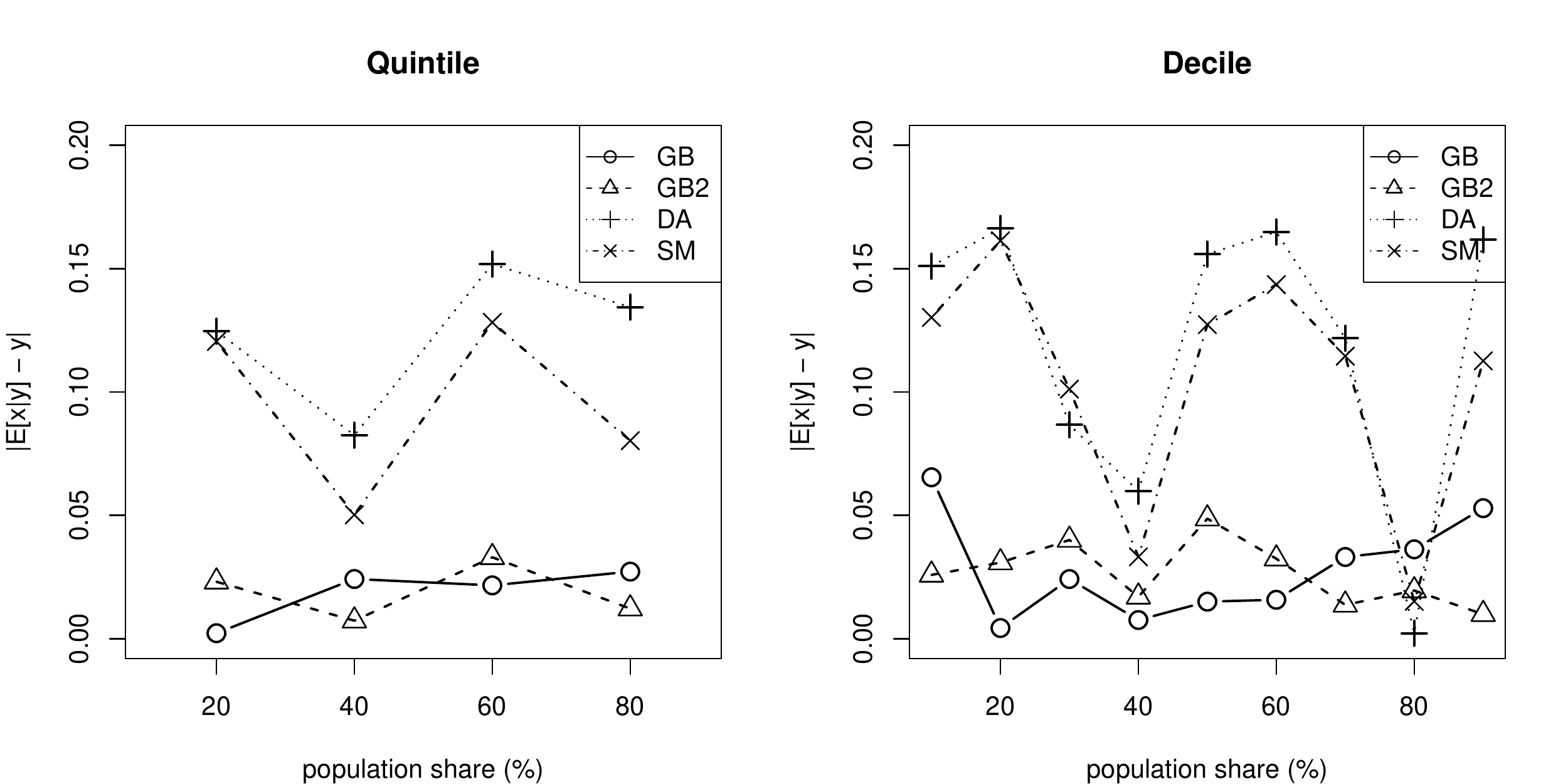}
\caption{Plots of $|E[x_j|\vy]-y_j|$ for the quintile and decile data}
\label{fig:real_2}
\end{figure}

Table~\ref{tab:real} presents the posterior means and 95\% credible intervals under the target tolerances.
For GB, we are able to learn about $a$ and $c$, while little learning about $p$ and $q$ occurred, similar to Setting~(i) of the simulation study. 
Similarly, for GB2, some learning about $a$ and $p$ occurred similar to Setting~(ii) of the simulation study. 
We can still obtain some insights on the shape of the underlying income distribution.
Figure~\ref{fig:real_2} presents the implied income distributions which are obtained by generating the random numbers from each distribution with the parameters fixed to their posterior means and scaling them with the theoretical standard deviations under these parameter values. 
The distribution shapes of GB and GB2 are almost identical. 
The figure also shows that DA and SM have higher density in the low income region and the right tails decays more quickly compared to GB and GB2.

\begin{table}[H]
\centering
\caption{Posterior summary for the FIES data}\label{tab:real}
{\scriptsize
\begin{tabular}{lcrrrrrrrrrrrr}\bottomrule
&& \multicolumn{3}{c}{GB} & \multicolumn{3}{c}{GB2}& \multicolumn{3}{c}{DA} & \multicolumn{3}{c}{SM} \\\cmidrule(lr){3-5}\cmidrule(lr){6-8}\cmidrule(lr){9-11}\cmidrule(lr){12-14}
Data&Parameter & Mean & \multicolumn{2}{c}{95\% CI}& Mean & \multicolumn{2}{c}{95\% CI} & Mean & \multicolumn{2}{c}{95\% CI} & Mean & \multicolumn{2}{c}{95\% CI} \\\hline
Quintile&$a$& 1.9334 & ( 1.2888, &  2.9330 ) & 2.0548 & ( 1.3746, & 3.1105 ) & 4.2602 & ( 4.0546, & 4.4624 ) & 3.6679 & ( 3.4981, & 3.8495 ) \\
        &$c$& 0.9686 & ( 0.9308, &  0.9974 ) &        &           &           &        &           &          &        &           &          \\
        &$p$& 3.5721 & ( 1.4269, &  7.0048 ) & 2.8919 & ( 1.2766, & 5.5099 ) & 0.8210 & ( 0.7371, & 0.9264 ) &        &           &          \\
        &$q$& 2.9111 & ( 1.2464, &  5.6645 ) & 3.4733 & ( 1.5834, & 6.5084 ) &        &           &          & 1.2568 & ( 1.1043, & 1.4290 ) \\
        &$G$& 0.2459 & ( 0.2414, &  0.2504 ) & 0.2464 & ( 0.2420, & 0.2509 ) & 0.2482 & ( 0.2432, & 0.2538 ) & 0.2474 & ( 0.2431, & 0.2519 ) \\\hline
Decile  &$a$& 1.9594 & ( 1.3073, &  2.9622 ) & 1.9800 & ( 1.3841, & 2.7915 ) & 4.2962 & ( 4.1005, & 4.5123 ) & 3.5924 & ( 3.4697, & 3.7289 ) \\
        &$c$& 0.9762 & ( 0.9470, &  0.9982 ) &        &           &           &        &           &          &        &           &          \\
        &$p$& 3.4502 & ( 1.4621, &  6.7507 ) & 3.0384 & ( 1.4702, & 5.6317 ) & 0.7938 & ( 0.7141, & 0.8853 ) &        &           &          \\
        &$q$& 3.0114 & ( 1.3057, &  5.6468 ) & 3.5858 & ( 1.9112, & 6.3023 ) &        &           &          & 1.3302 & ( 1.2005, & 1.4797 ) \\
        &$G$& 0.2461 & ( 0.2418, &  0.2503 ) & 0.2465 & ( 0.2420, & 0.2508 ) & 0.2486 & ( 0.2434, & 0.2544 ) & 0.2470 & ( 0.2427, & 0.2513 ) \\\bottomrule
\end{tabular}
}
\end{table}

\begin{figure}[H]
\centering
\includegraphics[width=\textwidth]{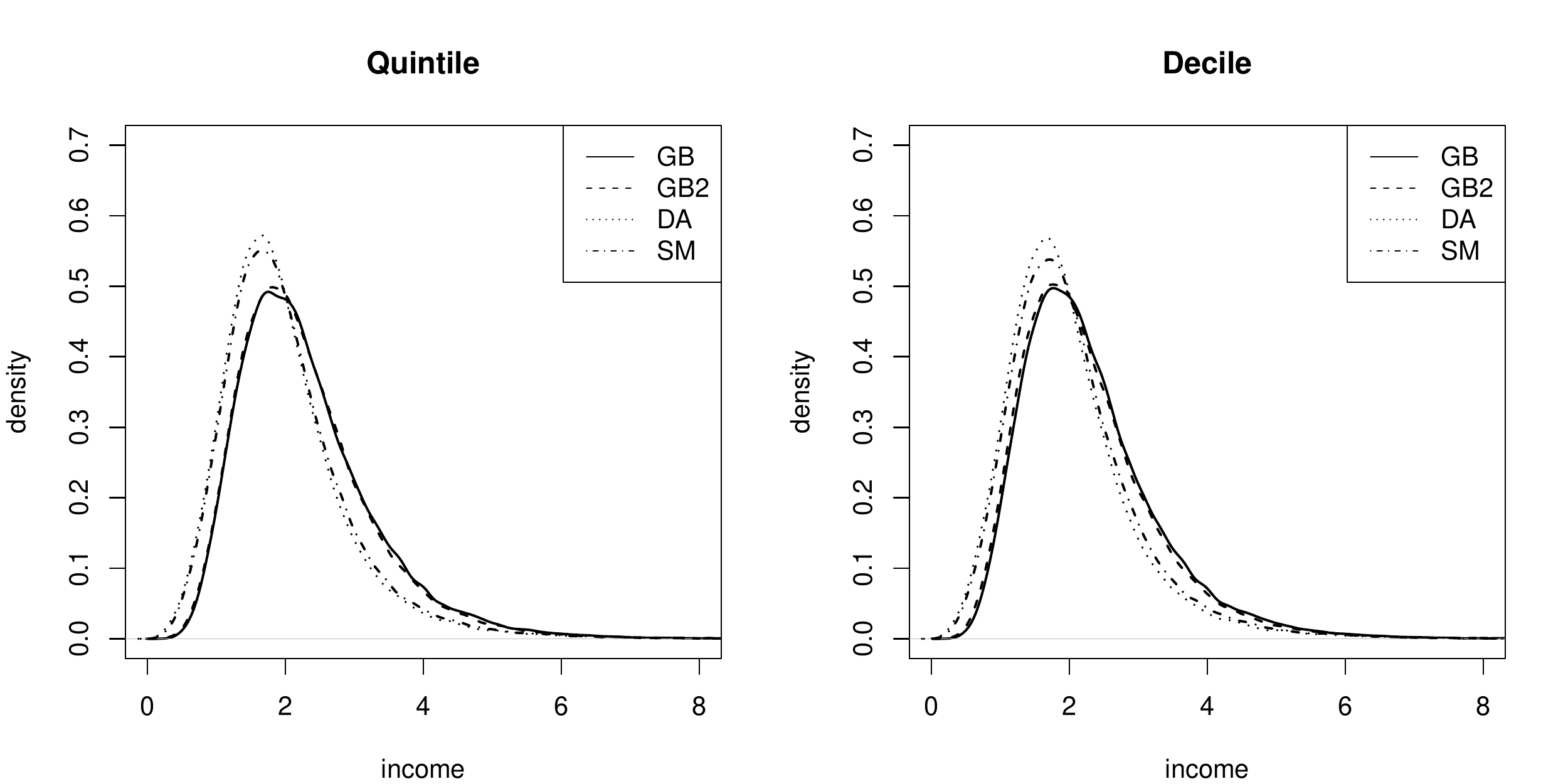}
\caption{Implied income distributions for the quintile and decile data}
\label{fig:real_1}
\end{figure}

The goodness of fit of the income models can be quantified through the marginal likelihood, which is calculated following Didelot~\etal~(2011). 
The log marginal likelihoods for GB, GB2, DA, and SM are, respectively, $-3.971$, $-2.110$, $-8.037$, and $-5.675$ for the quintile data and $-4.064$, $-2.100$, $-8.736$, and  $-5.427$ for the decile data. 
Based on the marginal likelihoods, GB2 is supported the most by the both data followed by GB. 
This result is consistent with McDonald and Xu~(1995) and is also in line with the argument made by Kleiber and Kotz~(2003). 
The goodness of fit can be also checked through the simulating function by plotting the absolute difference between the posterior mean of the simulated income share $x_j$ and the observed income share $y_j$, $|E[x_j|\vy]-y_j|$, for $j=1,\dots,k-1$ under each model. 
Figure~\ref{fig:real_2} shows that the absolute differences under GB and GB2 are generally smaller than those under DA and SM for both quintile and decile data, also suggesting the use of a more flexible class of income distributions. 

The posterior distributions of the Gini coefficient are compared with the nonparametric bounds of Gastwirth~(1972), in which a Gini estimate should be included.
The nonparametric bounds are given by $(0.2310,0.2545)$ and $(0.2419,0.2484)$ for the quintile and decile data, respectively.
This can be seen from Figure~\ref{fig:real_2}, which presents the posterior distributions of the Gini coefficient.
The shaded area in the figure represents the region inside the nonparametric bounds.
The marks on the horizontal axis represent the posterior means.
For the quintile data, all models resulted the posterior distributions of the Gini coefficient which are fairly concentrated within the nonparametric bounds. 
The posterior probabilities that the Gini coefficient is included in the bounds are $1.000$, $0.999$, $0.984$, and $0.998$ for GB, GB2, DA, and SM, respectively. 
In the case of the decile data, the figure shows that the bodies of the posterior distributions under GB, GB2, and SM are included in the nonparametric bounds. 
Only the left half of the posterior distribution is include in the bounds under DA and the posterior mean is outside the bounds. 
The posterior probabilities of the Gini coefficient included inside the bounds are $0.818$, $0.776$, $0.468$, and $0.718$ for GB, GB2, DA, and SM, respectively. 
While this result indicates the limitation that the posterior distribution obtained by using the proposed method does not shrink as fast as the nonparametric bounds, it is consistent with the results of the simulation studies. 
Nonetheless, GB2 appear to be the most appropriate income model among the four in terms of goodness of fit and the Gini coefficient. 

For comparison purpose, Figure~\ref{fig:real_2} also presents the posterior distributions of the Gini coefficient from SOS. 
For the quintile data, the posterior distributions appear to be more dispersed and scattered across regions. 
For the decile data, GB and GB2 produced the posterior distributions concentrated around the bounds with the posterior probabilities given by $0.710$ and $0.656$, respectively. 
Contrary, the posterior distributions under DA and SM are located away from the bounds. 
Therefore, the proposed ABC method  also provides more reliable estimates of Gini coefficient in terms of the nonparametric bounds.

\begin{figure}[H]
\centering
\includegraphics[width=\textwidth]{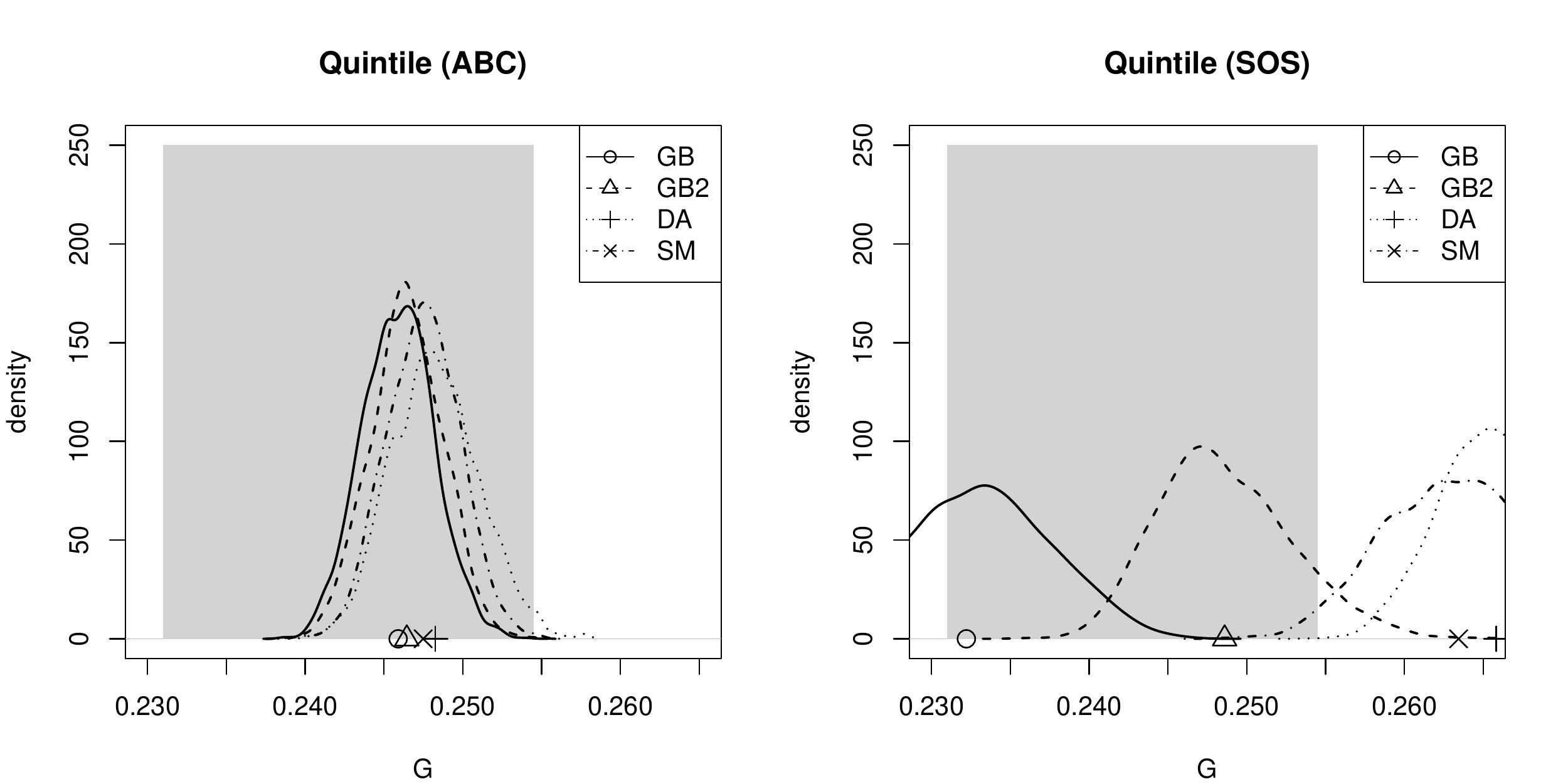}\\
\includegraphics[width=\textwidth]{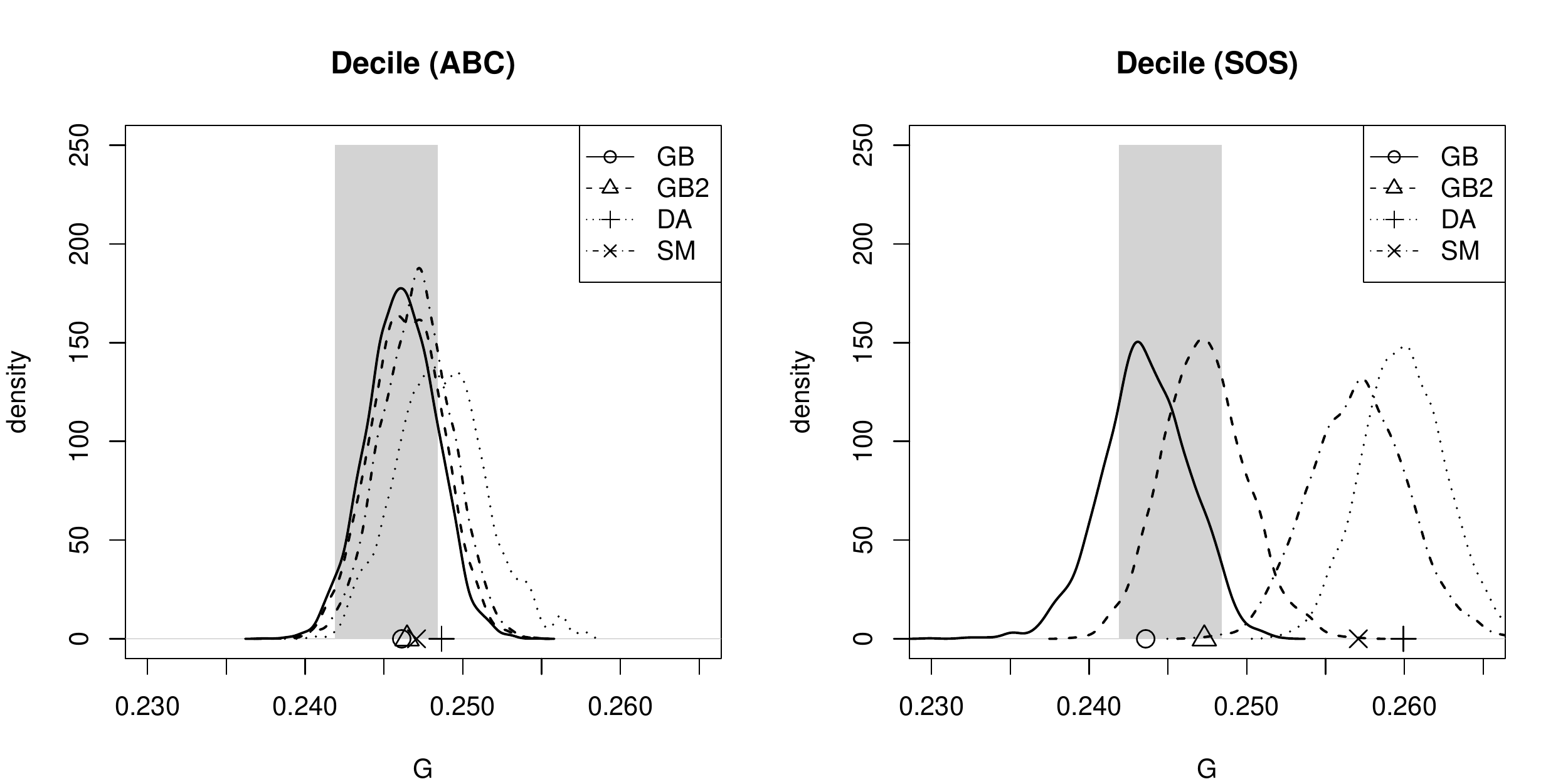}
\caption{Posterior distributions of the Gini coefficient using the proposed ABC approach and the MCMC approach using the level income for the quintile and decile data}
\label{fig:real_3}
\end{figure}

\section{Discussion}\label{sec:conc}
We proposed a new Bayesian approach to estimate the Gini coefficient from the Lorenz curve of a hypothetical income distribution based on grouped data and the ABC method based on the SMC algorithm is adopted for parameter estimation.
From the simulation study, the proposed approach is found to perform comparably with or better than the existing methods with respect to the Gini coefficient estimation.
Our approach is found to be particularly valuable in the where the number of group is small as in quintile data.
In the application to the Japanese data, the usefulness of the proposed approach assuming the class of GB distribution is illustrated by showing that the posterior distributions of the Gini coefficient are included within the nonparametric bounds with relatively high posterior probabilities and by presenting the income distributions implied from the hypothetical distributions.

Further, the numerical examples presented in this paper illuminate the limitation of the present study.
Some parameters of the hypothetical distribution may not be identified when the number of parameters is large as in the cases of GB and GB2, because the information contained in grouped data is severely limited.
Further, the posterior distribution of the Gini coefficient from the proposed approach does not shrink as fast as the nonparametric bounds as the number of income groups increases. 
Therefore, reconciling the goodness of fit and the accuracy of the Gini estimate  when we have more groups in the data provide a direction for the future research.

\end{document}